\documentstyle[12pt,twoside]{article}

\newcommand{\beqn}{\begin{eqnarray}}
\newcommand{\eeqn}{\end{eqnarray}}
\newcommand{\be}{\begin{equation}}
\newcommand{\ee}{\end{equation}}
\newcommand{\ba}{\begin{array}}
\newcommand{\ea}{\end{array}}
\newcommand{\pa}{\partial}
\newcommand{\re}{\ref}
\newcommand{\ci}{\cite}
\newcommand{\la}{\label}
\newcommand{\bfr}{\begin{flushright}}
\newcommand{\efr}{\end{flushright}}
\newcommand{\bfl}{\begin{flushleft}}
\newcommand{\efl}{\end{flushleft}}
\newcommand{\fr}{\frac}
\newcommand{\ov}{\overline}

\newcommand{\ve}{\varepsilon}
\newcommand{\de}{\delta}
\newcommand{\al}{\alpha}
\newcommand{\La}{\Lambda}
\newcommand{\om}{\omega}
\newcommand{\Om}{\Omega}
\newcommand{\si}{\sigma}
\newcommand{\Si}{\Sigma}
     \newcommand{\ds}{\displaystyle}
\newcommand{\pr}{\prime}
\newcommand{\De}{\Delta}

\newcommand{\na}{\nabla}
\newcommand{\br}{|\kern-.25em|\kern-.25em|}
\newcommand{\brr}{{|\kern-.15em|\kern-.15em|\kern-.15em}\,}
\newcommand{\ddd}{\st{.\kern-.07em.\kern-.07em.}}

      \def\N{{\rm I\kern-.1567em N}}                              
\def\R{{\rm I\kern-.1567em R}}                              
\def\C{{\rm C\kern-4.7pt                                    
\vrule height 7.7pt width 0.4pt depth -0.5pt \phantom {.}}}
\def\Z{{\sf Z\kern-4.5pt Z}}                                

\newcommand{\supp}{\mathop{\rm supp}\nolimits}

\newtheorem{theorem}{Theorem}[section]

\newtheorem{definition}[theorem]{Definition}

\newtheorem{lemma}[theorem]{Lemma}

\newtheorem{remark}[theorem]{Remark}

\newtheorem{cor}[theorem]{Corollary}
\newtheorem{pro}[theorem]{Proposition}

\begin{document}
\begin{titlepage}
\hspace{6cm} {\em Journal of Statistical Physics} {\bf 108} (2002), no.4, 1219-1253
\bigskip\bigskip\bigskip
\vspace{2cm}

{\it Dedicated to Ya.G.~Sinai on the occasion of his $65$'th anniversary}
\vspace{0mm}

\begin{center}
{\Large\bf  On Convergence to  Equilibrium Distribution, II.\bigskip\\
The Wave  Equation in Odd Dimensions, with Mixing}
\end{center}\vspace{0mm}
{\bf T.V.~Dudnikova
 \footnote{Supported partly by research grants of
  DFG (436 RUS 113/615/0-1) and
 RFBR (01-01-04002).}
}
Mathematics Department,
 Elektrostal Polytechnical Institute,\\
        Elektrostal, 144000 Russia;
e-mail:~misis@elsite.ru.
 \medskip\\
 {\bf A.I.~Komech
 \footnote{Supported partly by the
Institute of Physics and Mathematics of Michoacan in Morelia,
the Max-Planck Institute for the Mathematics in Sciences
(Leipzig) and by
research grant of
  DFG (436 RUS 113/615/0-1).}
 }
 Mechanics and Mathematics Department, Moscow State University,
 Moscow, 119899 Russia;
 e-mail:~komech@mech.math.msu.ru.
 \medskip\\
{\bf N.E.Ratanov
 \footnote{Supported partly by
research grants of
RFBR (99-01-00989).}
}
 Economics  Department, Chelyabinsk State University,
Chelyabinsk, 454021 Russia;
e-mail:~nickita@cgu.chel.su.
\medskip\\
 {\bf Yu.M.~Suhov}
 Statistical Laboratory,
Department of Pure Mathematics and Mathematical Statistics,
University of Cambridge,
 Cambridge, UK;
 e-mail:~Y.M.Suhov@statslab.cam.ac.uk.
\vspace{5mm}

\begin{abstract}
The paper considers the wave equation, with constant 
or variable coefficients in $\R^n$,
with odd $n\geq 3$. We study the asymptotics of 
the distribution $\mu_t$
of the {\it random solution} at  time  $t\in\R$ as $t\to\infty$.
It is assumed that the initial measure $\mu_0$
has zero mean, translation-invariant covariance matrices,
and finite expected energy density. We also assume that 
$\mu_0$ satisfies a Rosenblatt- or
Ibragimov--Linnik-type space mixing condition.
The main result is the convergence of $\mu_t$ to
a Gaussian measure $\mu_\infty$ as $t\to\infty$, 
which gives a Central Limit Theorem (CLT) for the wave equation.
The proof for the case of constant coefficients is based on
an analysis of long-time asymptotics
of the solution in the Fourier representation
and  Bernstein's `room-corridor'  argument.
The case of variable coefficients is treated 
by using a version of the scattering theory for infinite
energy solutions, based on Vainberg's results on local energy decay.

\end{abstract}
\end{titlepage}

 \section{Introduction}

This paper can be considered as a continuation of \ci{DKKS}.
Here we develop a probabilistic analysis for the
linear wave equation (WE) in $\R^n$, with odd $n\geq 3$:
\beqn
\left\{\ba{l}\ddot u(x,\,t)  =   
Lu(x,t)\equiv \sum\limits_{i,j=1}^n 
\pa_i(a_{ij}(x)\pa_j u(x,\,t)) 
- a_0(x)\,u(x,\,t),\\ 
u|_{t=0} = u_0(x),~~\dot u|_{t=0} = v_0(x), 
\ea \right.\la{1.1}\eeqn 
where $\ds\pa_i\equiv \fr\pa{\pa x_i}$, $x\in\R^n$, $t\in\R$. 
We assume that the coefficients of  
equation (\re{1.1}) are constant outside a bounded region, 
more precisely, $a_{ij}(x)=\de_{ij}$ for $|x|\geq$ const. 
Moreover, we assume that a {\it nontrapping} condition is satisfied, 
i.e. all rays of a geometrical optics associated with (\re{1.1}) 
go to infinity (see Condition {\bf E3} in Section 2.1). 
Denote $Y(t)=(Y^0(t),Y^1(t))\equiv (u(\cdot,t),\dot u(\cdot,t))$, 
$Y_0=(Y^0_0,Y^1_0)\equiv (u_0,v_0)$. Then  (\ref{1.1}) becomes 
\be\la{1'}\dot Y(t)={\cal A}Y(t),\,\,\,t\in\R;\,\,\,\,Y(0)=Y_0.\ee  
Here we set: 
\be\la{A}{\cal A}=\left(\begin{array}{cc} 
0 & 1\\ A & 0 \end{array}\right),\ee 
where  $A=\sum\limits_{i,j=1}^n\pa_i (a_{ij}(x) \pa_j)- a_0(x)$. 
We suppose that the initial date $Y_0$ is a random function with 
zero mean living in a functional phase space ${\cal H}$ representing 
states with finite local energy; the distribution of $Y_0$ is 
denoted by $\mu_0$. Denote by  $\mu_t$, $t\in\R$, the measure on  
${\cal H}$ giving the distribution of the random solution  $Y(t)$ 
to problem (\re{1'}). We assume that the initial covariance 
functions (CFs) are translation-invariant, i.e. 
\be\la{1.9'} 
Q^{ij}_0(x,y):= E\Big( Y_0^i(x)Y_0^j(y)\Big) = 
q^{ij}_0(x-y),\,\,\,x,y\in\R^n.\;\;\;i,j=0,1.\ee 
Next, we assume that the initial `mean energy density'   
is finite: 
\be\la{med} 
e_0:=E \Big(\vert v_0(x)\vert^2+\vert \nabla u_0(x)\vert^2 
+ \vert u_0(x)\vert^2\Big)=q_0^{11}(0) 
-\De q_0^{00}(0)+q_0^{00}(0)<\infty,\,\,\,\,x\in\R^n.\ee 
Finally, it is assumed that $\mu_0$ satisfies a space-mixing 
condition. Roughly speaking, it means that 
\be\la{mix} Y_0(x)\,\,\,\,   and \, \, \,\,Y_0(y) 
\,\,\,\,  are\,\,\,\, asymptotically\,\,\,\, independent\,\, \,\, 
 as \,\, \,\, |x-y|\to\infty. \ee 
Our main result establishes the convergence 
\be\la{1.8i} 
\mu_t \rightharpoondown 
\mu_\infty,\,\,\,\, t\to \infty. 
\ee 
to a stationary measure $\mu_\infty$, 
that is Gaussian and supported in space ${\cal H}$. 
\medskip


Previously, results of this kind have been obtained in  
\ci{KM, R1, R2,R3}, for translation-invariant initial 
measures $\mu_0$. However, the detailed proofs were
not published because of their length. Another
drawback was the absence of a 
unifying argument indicating the limits of the methods. 
In this paper, such an argument is presented, which also 
improves the assumptions and makes the proofs much shorter. 
Like  \ci{DKKS}, the argument is based on a systematic
use of Fourier transform (FT) and a duality argument 
in Lemma \re{ldu} (cf. works \ci{BPT,SS,SL} concerning 
FT arguments for lattice systems). This is used in 
conjunction with the {\it strong} Huyghen's
principle for the WE and the fact that the rank of the
Hessian of the restricted dispersion relation (\re{Hess}) 
equals $n-1$.
We also found a more efficient method to estimate  
higher order momentum functions and to complete some 
details in the proof of scattering theory results for 
the case of variable coefficients. The mixing condition  
has been  used  in \ci{BDS, BPT, SS,SL} to prove the  
convergence for various classes of systems. In this paper 
it is used in the context of the WE.  
\medskip 
 
We prove relation (\re{1.8i}) by using the strategy
similar to \ci{DKKS}. At first, we prove (\re{1.8i}) for 
the equations with constant coefficients 
$a_{ij}(x)\equiv\de_{ij}$, in three steps. 
\\ 
{\bf~~ I.} 
The family of measures 
 $\mu_t$, $t\geq 0$, is compact in an appropriate Fr\'echet space.\\ 
{\bf~ II.} 
The CFs converge to a limit: for $i,j=0,1$, 
 \be\la{corf} 
Q_t^{ij}(x,y)=\int Y^i(x)Y^j(y)\mu_t(dY) 
\to Q_\infty^{ij}(x,y),\,\,\,\,t\to\infty. 
\ee 
{\bf III.} 
The characteristic functionals converge to Gaussian: 
\be\la{2.6i} 
 \hat\mu_t(\Psi ) 
 =   \int \exp({i\langle Y,\Psi\rangle })\mu_t(dY) 
\rightarrow \exp\{-\fr{1}{2}{\cal Q}_\infty( \Psi, \Psi)\}, 
\,\,\,\,t\to\infty, 
 \ee 
where  
$\Psi$ is an arbitrary element of a dual space  and 
${\cal Q}_\infty$ is the quadratic form 
with the integral kernel 
$(Q^{ij}_\infty(x,y))_{i,j=0,1}$. 
\medskip 
 
Property {\bf I} follows from Prokhorov's  
Compactness Theorem with the help of arguments 
from \ci{VF}. First, one proves a uniform bound for the 
mean local energy in measure $\mu_t$ with the help of 
the FT. The conditions of Prokhorov' Theorem the  follow 
from  Sobolev's Embedding Theorem. Property {\bf II} is 
deduced from an analysis of oscillatory integrals 
arising in the FT. An important role is attributed to  
Proposition \re{l40.1} which establishes useful properties 
of the CFs in the FT deduced from the mixing condition. 
 
On the other hand, the FT alone is not sufficient  
to prove property {\bf III} even in the case of constant 
coefficients. The reason is that a function of infinite 
energy gives a singular generalised function in the FT, 
and an exact interpretation of condition (\re{mix}) in 
these terms is unclear. We deduce  property  {\bf III}  
from a representation for the solution in the coordinate space, 
which manifests a dispersion of  waves.  In particular, for 
the case $n=3$ and $u_0(x)\equiv 0$,  Kirchhoff's formula holds: 
\be\la{Kir}u(x,t)= \frac{1}{4\pi t}\int\limits_{S_t(x)} 
v_0(y)  dS(y),\,\,\,\, x\in\R^3,\ee 
where $dS(y)$ is the Lebesgue measure on the sphere 
$S_t(x):\,|y-x|=t$. Then the proof of (\re{2.6i}) 
proceeds with a modification of Bernstein's `room-corridor' 
method, well-known in the random processes theory. Namely,  
we divide the sphere of integration in (\re{Kir}) into  
`rooms' $R^k_t,\,1\leq k\leq N$, of a fixed width $d\gg 1$, 
separated by `corridors' $C^k_t$ of a fixed width $\rho\ll d$. 
As the area $| R^k_t| \sim d^3=$const, the number 
$N\sim t^2$, and  (\re{Kir}) becomes 
\be\la{4.2} 
u(x,\,t) \sim \fr{\sum_{k=1}^N\,r^k_t}{\sqrt N},\ee 
where $r^k_t$ is the integral over $R^k_t$. 
The contribution of the `corridors' turns out to be 
negligible.   
Assume for a moment that $v_0(x)$ and $v_0(y)$ 
are independent for $|x-y|\ge \rho_0$. Then 
$r^k_t$ are independent if $\rho>\rho_0$, and 
random variable $u(x,t)$ is asymptotically 
Gaussian by the CLT. 
 
So, the CLT emerges from (\re{Kir}) because of 
integration over the sphere $|x'-x|=t$ 
and the first power of $t$ in the denominator. 
A similar geometrical structure 
of an integral over the sphere $|x'-x|=t$ 
emerges from Herglotz-Petrovskii's formulas  
in a general odd dimension $n\geq 5$. 
However, the extension of the argument based on 
(\re{Kir}), (\re{4.2}) is not straightforward for  
$n\geq 5$ as the Herglotz-Petrovskii's formulas  
contain  high-order derivatives of initial functions. 
 
We cover all odd values $n\ge 3$ in a  
unified  techniques by modifying  the approach developed in 
\ci{DKKS} for  the Klein-Gordon equation (KGE). However, 
for the KGE, the solution  is an  integral over the ball 
$|x'-x|\leq t$. This fact allowed us to use for the KGE
a rather different approach based on the analysis 
of an oscillatory integral where the phase function 
(`dispersion relation') has a nondegenerate Hessian.  
For the WE, the Hessian is degenerate, which requires 
additional constructions. Here we use the fact that 
the `restricted' Hessian has a maximal rank $n-1$, 
see (\re{Hess}). This leads to a weaker dispersion of 
waves  comparing to the KGE. Newertheless, we still 
obtain the representation of the solution as a sum of 
weakly dependent random variables. Then (\ref{2.6i}) 
follows from the CLT. However, checking  
the Lindeberg condition for the WE requires some delicate 
calculations. Here, 
the deficiancy in dispersive properties is compensated 
by the reduction in dimension of the domain of integration 
due to the strong Huyghen's principle.  
 
All three steps {\bf I}-{\bf III}  
of our argument  rely on the mixing 
condition. Simple examples show that  the 
convergence to a Gaussian measure 
may fail when the mixing condition fails: 
if we take $u_0(x)\equiv 0$ and 
$v_0(x)\equiv\pm 1$ with probability $p_\pm=0.5$, 
then $u(x,\,t)\equiv\pm t$ almost sure. 
 
Finally, we prove the convergence in (\re{1.8i}) 
for  problem (\re{1.1}) with  variable coefficients. 
In this case explicit formulas for the solution  
are unavailable. To prove (\re{1.8i}) in this case, 
we use a version of the scattering theory for solutions of 
infinite global energy (this strategy is similar to \ci{BM}). 
This allows us to reduce the proof to the case 
of constant coefficients. Namely, we establish  the 
long-time asymptotics 
\be\la{lt} 
U(t)Y_0=\Theta U_0(t)Y_0+\rho(t)Y_0,\,\,\,t>0, \ee 
where $U(t)$ is the dynamical group of  
Eqn (\re{1.1}),  $U_0(t)$ corresponds to 
the constant coefficients $a_{jk}(x)\equiv\de_{ij}$, 
and $\Theta$ is a `scattering operator'. The remainder, 
$\rho(t)$, is small in local energy seminorms 
$\Vert\cdot\Vert_R$, $\forall R>0$: 
\be\la{rema} 
\Vert\rho(t)Y_0\Vert_R\to 0,\,\,\,t\to\infty. \ee 
The scattering theory results are based on the Vainberg's 
estimates for the local energy decay; see \cite{V89}. 
\begin{remark}\la{scat}  
\vspace{-2mm} 
{\rm  
i) Under our assumptions on initial  
measure $\mu_0$, initial date $Y_0$  
has an infinite energy. Therefore, 
the standard scattering theory, for the solutions of 
a finite energy (see, e.g., \ci{LP}), is not sufficient  
for our purposes.\\ 
ii) The order of the operators  in product    
$\Theta U_0(t)$ in (\re{lt}) 
differs from  that in 
$U_0(t)\Theta$ considered in 
the scattering theory  
of  finite energy solutions.  
An  asymptotics with the order $U_0(t)\Theta$ would  
mean that  $Y(t)$ is close to a solution  
of the unperturbed equation. This is impossible  
 for the solutions of infinite energy  
as they do not converge locally to zero,  
hence the perturbation terms in the equation  
are not negligible.}  
\end{remark} 
 
\vspace{-2mm} 
The paper is organised as follows. 
In Section 2 
we formally state our main result. 
Sections 3-7 deal with the case of constant coefficients: 
main results are stated in Section 3, 
the compactness (Property {\bf I}) and the convergence  
(\re{corf}) are proved in Section 4, and convergence  
(\re{2.6i})  in Sections 5,6. In Section 7 we check the Lindeberg condition. 
In Section 8 we construct the scattering theory, and 
in Section 9 establish convergence (\re{1.8i}) for variable coefficients. 
Section 10  discusses Vainberg's estimates. 
In Appendix we collected the  FT calculations. 
\bigskip\\ 
{\bf Acknowledgements} 
Authors thank 
 V.I.Arnold, A.Bensoussan, I.A.Ibragimov, 
H.P.McKean,  J.Lebowitz, A.I.Shnirelman, 
 H.Spohn, B.R.Vainberg and M.I.Vishik  for 
fruitful discussions and remarks.

\section{Main results} 
\setcounter{equation}{0} 
\subsection{The notation} 
 
Denote by $D$ the space of real functions $C_0^\infty (\R^n)$. 
We assume that 
the following properties {\bf E1--E3} of  Eqn (\re{1.1}) 
are satisfied: 
\medskip\\ 
{\bf E1} $a_{ij}(x) = \delta _{ij} + b_{ij}(x)$, 
 where $b_{ij}(x) \in D$; also $a_0(x)\in D$. 
\medskip\\ 
{\bf E2} $a_0(x)\geq 0$, and the 
hyperbolicity condition holds: $\exists\al>0$ 
\be\la{1.3} H(x, k )\equiv 
\fr 12 \sum_{i,j=1}^n a_{ij}(x) k_i  k_j\geq \al | k|^2, 
\,\,\,\,x, k \in \R^n. \ee 
{\bf E3} A non-trapping condition \ci{V89}: 
for $(x(0), k(0))\in\R^n\times\R^n$ with $ k(0)\neq 0$, 
\be\la{1.4}\vert x(t)\vert \rightarrow \infty\quad
\rm{as}\quad t\rightarrow \infty,\ee 
where $(x(t), k(t))$ is a solution to the Hamiltonian system 
$$ \dot x(t)=\,\,\na_k H(x(t), k(t)),~~ 
\dot  k(t)=-\na_x H(x(t), k(t)).$$ 
{\bf Example}. {\bf E1}-{\bf E3} hold for the acoustic equation 
with constant coefficients 
\be \ddot u(x,\,t) = \Delta u(x,\,t),\,\,\,\,\, x\in \R^n. 
\la{2.1'} \ee 
For instance, {\bf E3} follows because $\dot k(t)\equiv 
0\Rightarrow x(t)\equiv k(0)t + x(0)$. 
\medskip 
 
We assume that the initial date $Y_0$ 
belongs to the phase space ${\cal H}$ defined below. 
\begin{definition} \la{d1.1} 
${\cal H}\equiv H_{\rm loc}^1(\R^n)\oplus H_{\rm loc}^0(\R^n)$ 
is the Fr\'echet space of pairs $Y(x)\equiv(u(x),v(x))$ 
of real  functions $u(x)$, $v(x)$, endowed with local 
energy seminorms 
\beqn \la{1.5} 
\Vert Y\Vert^2_{R}= \int\limits_{|x|<R} 
\Big(|v(x)|^2+|\nabla u(x)|^2+|u(x)|^2\Big) dx<\infty, 
~~\forall R>0. \eeqn 
\end{definition} 
 
Proposition \re{p1.1} follows from \ci[Thms V.3.1, V.3.2]{Mikh} 
as the speed of propagation for Eqn (\re{1.1}) is finite. 
 
\begin{pro}    \la{p1.1} 
i) For any  $Y_0 \in {\cal H}$ 
there exists  a unique solution 
$Y(t)\in C(\R, {\cal H})$ 
to Cauchy problem (\re{1'}).\\ 
ii) For any  $t\in \R$, the operator $U(t):Y_0\mapsto  Y(t)$ 
is continuous in ${\cal H}$. 
\end{pro} 
 
We now introduce appropriate Hilbert spaces 
of initial data of infinite 
energy. Let $\de$ be an arbitrary positive number. 
\begin{definition}\la{d6.1} ${\cal H}_{\delta}$ 
is the  Hilbert space of the functions $Y=(u,v)\in  {\cal H}$ 
with a finite norm 
$$ \brr Y\brr^2_{\delta}=\int e^{-2\delta|x|} 
\Big(|v(x)|^2+|\nabla u(x)|^2+|u(x)|^2\Big)\,dx<\infty. $$ 
\end{definition} 
 

Let us choose a function $\zeta(x)\in C_0^\infty(\R^n)$ with 
$\zeta(0)\ne 0$. Denote by $H^s_{\rm loc}(\R^n),$ $s\in\R$,  
the local Sobolev spaces, i.e. the Fr\'echet spaces 
of distributions $u\in D'(\R^n)$ with finite seminorms 
$$\Vert u\Vert _{s,R}:= \Vert\La^s\Big(\zeta(x/R)u\Big)
\Vert_{L^2(\R^n)},$$ 
where $\La^s v:=F^{-1}_{k\to x}(\langle k\rangle^s\hat v(k))$, 
$\langle k\rangle:=\sqrt{|k|^2+1}$, and $\hat v:=F v$ is the FT 
of a tempered distribution $v$. For $\psi\in D$ define 
$F\psi ( k)= \ds\int e^{i k\cdot x} \psi(x) dx$.  
 
\begin{definition}\la{d1.2} 
For $s\in\R$  denote 
${\cal H}^{s}\equiv H_{\rm loc}^{1+s }(\R^n) 
\oplus H_{\rm loc}^{s }(\R^n)$. 
\end{definition} 
 
Using standard techniques of pseudodifferential operators  
and Sobolev's Embedding Theorem (see, e.g. \ci{H3}), it is 
possible to prove that ${\cal H}^0={\cal H}\subset 
{\cal H}^{-\ve }$ for every $\ve>0$, and the embedding  
is compact. We denote by $\langle \cdot ,\cdot \rangle $ 
scalar product in real Hilbert space $L^2(\R^n)$ or in 
$L^2(\R^n)\otimes\R^N$ or its various extensions. 
 
\subsection{Random solution. Convergence to equilibrium} 
 
Let $(\Om,\Si,P)$ be a probability space 
with  expectation $E$ 
and ${\cal B}({\cal H})$ denote the Borel $\si$-algebra 
in ${\cal H}$. 
We assume that $Y_0=Y_0(\om,x)$ in (\re{1'}) 
is a measurable 
random function 
with values in $({\cal H},\,{\cal B}({\cal H}))$. 
In other words, $(\om,x)\mapsto Y_0(\om,x)$ is a measurable 
 map 
$\Om\times\R^n\to\R^2$ with respect to the  
(completed) 
$\si$-algebras 
$\Si\times{\cal B}(\R^n)$ and ${\cal B}(\R^2)$. 
Then 
$Y(t)=U(t) Y_0$ is also a measurable  random function 
with values in 
$({\cal H},{\cal B}({\cal H}))$ owing  to Proposition \re{p1.1}. 
We denote by $\mu_0(dY_0)$ a Borel probability measure  
in ${\cal H}$ 
giving 
the distribution of the  $Y_0$. 
Without loss of generality, 
 we assume $(\Om,\Si,P)= 
({\cal H},{\cal B}({\cal H}),\mu_0)$ 
and $Y_0(\om,x)=\om(x)$ for 
$\mu_0(d\om)\times dx$-almost all  
$(\om,x)\in{\cal H}\times\R^n$. 
 
\begin{definition} 
$\mu_t$ is a Borel probability measure in ${\cal H}$ 
which gives 
the distribution of $Y(t)$: 
\begin{eqnarray}\la{1.6} 
\mu_t(B) = \mu_0(U(-t)B),\,\,\,\, 
\forall B\in {\cal B}({\cal H}), 
\,\,\,   t\in \R. 
\eeqn 
\end{definition} 
 
Our main goal is to derive 
 the convergence of the measures $\mu_t$ as $t\rightarrow \infty $. 
We establish the weak convergence of  $\mu_t$ 
in the Fr\'echet spaces ${\cal H}^{-\ve }$ with any  $\ve>0$: 
\be\la{1.8} 
\mu_t\,\buildrel {\hspace{2mm}{\cal H}^{-\ve }}\over 
{- \hspace{-2mm} \rightharpoondown } 
\, \mu_\infty 
\quad\rm{as}\quad t\to \infty, 
\ee 
where $\mu_\infty$ is a Borel probability measure in space 
${\cal H}$. 
  By definition,    this means 
 the convergence 
 \be\la{1.8'} 
 \int f(Y)\mu_t(dY)\rightarrow 
 \int f(Y)\mu_\infty(dY)\quad\rm{as}\quad t\to \infty 
 \ee 
 for any bounded continuous functional $f(Y)$ 
 in space ${\cal H}^{-\ve }$. 
 
\begin{definition} 
The CFs of measure $\mu_t$ are 
defined by 
\be\la{qd} 
Q_t^{ij}(x,y)\equiv E \Big(Y^i(x,t)Y^j(y,t)\Big),~~i,j= 0,1,~~~~ 
{\rm for~~almost~~all}~~~ 
x,y\in\R^n\times\R^n 
\ee 
if the expectations in the RHS are finite. 
\end{definition} 
 
We set ${\cal D}=D\oplus D$, and 
$ 
\langle Y,\Psi\rangle  
= 
\langle Y^0,\Psi^0\rangle +\langle Y^1,\Psi^1\rangle  
$ 
for 
$Y=(Y^0,Y^1)\in {\cal H}$, and $ 
\Psi=(\Psi^0,\Psi^1)\in  {\cal D}$. 
For a Borel probability  measure $\mu$ in the space ${\cal H}$ 
we denote by $\hat\mu$ 
the characteristic functional (the Fourier transform of $\mu$) 
$$ 
\hat \mu(\Psi )  \equiv  \int\exp(i\langle Y,\Psi \rangle )\,\mu(dY),\,\,\, 
 \Psi\in  {\cal D}. 
$$ 
A  measure $\mu$ is called Gaussian (with zero expectation) if 
its characteristic functional has the form 
$$ 
\ds\hat { \mu} (\Psi ) =  \ds \exp\{-\fr{1}{2} 
 {\cal Q}(\Psi , \Psi )\},\,\,\,\Psi \in {\cal D}, 
$$ 
where ${\cal Q}$ is a  real nonnegative quadratic form in ${\cal D}$. 
A measure $\mu$ is called 
translation-invariant if  
$$ 
\mu(T_h B)= \mu(B),\,\,\,\,\,\forall B\in{\cal B}({\cal H}),  
\,\,\,\, h\in\R^n, 
$$ 
where $T_h Y(x)= Y(x-h)$.

\subsection{Mixing condition} 
Let $O(r)$ denote the set of all pairs of open subsets 
${\cal A},\>{\cal B}\subset \R^n$ at distance 
$\rho ({\cal A},\,{\cal B})\geq r$ and  $\sigma ({\cal A})$ be 
the $\sigma $-algebra of the subsets in ${\cal H}$ generated by all 
linear functionals $Y\mapsto\, \langle Y,\Psi\rangle $, 
where  $\Psi\in  {\cal D}$ 
with $ \supp \Psi \subset {\cal A}$. 
We define the 
Ibragimov-Linnik mixing coefficient  
of a probability  measure  $\mu_0$ on ${\cal H}$ 
by (cf \ci[Dfn 17.2.2]{IL})  
\be\la{ilc} 
\varphi(r)\equiv 
\sup_{({\cal A},{\cal B})\in O(r)} \sup_{ 
\ba{c} A\in\si({\cal A}),B\in\si({\cal B})\\ \mu_0(B)>0\ea} 
\fr{| \mu_0(A\cap B) - \mu_0(A)\mu_0(B)|}{ \mu_0(B)}. 
\ee 
\begin{definition} 
 Measure $\mu_0$ satisfies the  {strong uniform} 
Ibragimov-Linnik mixing condition if 
\be\la{1.11} 
\varphi(r)\to 0,\quad r\to\infty. 
\ee 
\end{definition} 
Below, we  specify the rate of the decay.

\subsection{Main theorem} 
We assume that  measure $\mu_0$ 
satisfies the following properties {\bf S0--S3}: 
\bigskip\\ 
{\bf S0}  
$\mu_0$ has the zero expectation value, 
\be 
EY_0(x)  \equiv  0,\,\,\,\,\,\,\,x\in\R^n. 
\la{1.10'} 
\ee 
{\bf S1} The CFs 
of  $\mu_0$  are translation invariant, i.e. Eqn 
(\re{1.9'}) holds 
for almost all $x,y\in\R^n$. 
\\ 
{\bf S2} $\mu_0$ has a finite ``mean energy density'',  
i.e. Eqn (\re{med}) holds.\\ 
{\bf S3} Measure 
 $\mu_0$ satisfies the strong uniform 
Ibragimov-Linnik mixing condition, and 
\be\la{1.12} 
\ov\varphi\equiv 
\int\limits  
_0^\infty r^{n-2}\varphi^{1/2}(r)dr <\infty. 
\ee 
\bigskip

\begin{remark}\la{infen} 
{\rm (\re{med}) implies that 
  $\mu_0$ is concentrated in 
${\cal H}_\de$ for all $\delta>0$, since 
\be\la{bd} 
\int\brr Y_0\brr^2_{\delta}\,\mu_0(dY_0) 
=e_0\int \exp({-2\delta|x|})\,dx  <\infty. 
\ee 
} 
\end{remark} 
 
Let ${\cal E} (x)=-C_n |x|^{2-n}$ be the fundamental solution of the 
Laplacian, i.e. 
$\De {\cal E} (x) = \de(x)$ for $x\in \R^n$. 
Define, for almost all $x,y\in\R^n$, 
 the matrix-valued function 
 \be\la{Q} 
Q_\infty(x,y)= 
 \Bigl(Q_\infty ^{ij}(x, y)\Bigr)_{i,j= 0,1} = 
 \Bigl(q_\infty ^{ij}(x-y)\Bigr)_{i,j= 0,1}, 
\ee 
where 
 \beqn\la{1.13} 
\Bigl(q_\infty^{ij}\Bigr)_{i,j=0,1}= 
\frac{1}{2}\left( 
\ba{ll} 
q_0^{00}-{\cal E} * q_0^{11} & 
q_0^{01}-~~q_0^{10} \\ 
q_0^{10}-~~q_0^{01} & q_0^{11}-\De q_0^{00} 
\ea 
\right). 
\eeqn 
According to \ci[Lemma 17.2.3]{IL} (see Lemma \re{il} i) below), 
the derivatives 
$\pa_z^{\al}q_0^{ij}$ are bounded 
by  mixing coefficient: 
$\forall\al\in\Z_+^n$ with 
$|\al|\le 2-i-j$ (including $\al=0$), $i,j=0,1,$ 
\be\la{1.19} 
|\pa_z^{\al}q^{ij}_0(z)|\le 2 e_0\varphi^{1/2}(|z|), 
~~\forall z\in\R^n. 
\ee 
Hence,  (\re{1.12}) implies 
the existence of the convolution ${\cal E}*q_0^{11}$  
in (\re{1.13}).  
Denote by 
${\cal Q}_{\infty} (\Psi, {\Psi})$ a real quadratic form 
in  
${\cal D}$   defined by
\be\la{qpp}
{\cal Q}_\infty (\Psi, {\Psi})=\sum\limits_{i,j=0,1}~
\int\limits_{\R^n\times\R^n}
Q_{\infty}^{ij}(x,y)\Psi^i(x)\Psi^j(y)
dx~dy.
\ee 
Our main result 
is the following theorem. 
\medskip\\ 
{\bf Theorem A}~{\it 
Let $n\geq 3$ be odd, and  {\bf E1--E3},  
{\bf S0--S3} hold. 
Then \\ 
i) The convergence in (\re{1.8}) holds for 
any $\ve>0$.\\ 
ii) The limiting measure 
$ \mu_\infty $ is a Gaussian equilibrium 
measure on ${\cal H}$.\\ 
iii) The limiting characteristic functional has the form 
$$ 
\ds\hat { \mu}_\infty (\Psi ) = \exp 
\{-\fr{1}{2}  {\cal Q}_\infty (W\Psi ,W\Psi )\}, 
\,\,\, 
\Psi \in {\cal D}, 
$$ 
where $W:\,{\cal D}\rightarrow {\cal H}_\de^\pr$ 
is a linear continuous operator for sufficiently small  
$\de>0$, and  the quadratic form 
${\cal Q}_\infty$ is continuous in  
${\cal H}_\de^\pr$.}

 
\subsection{Remarks on various mixing conditions 
for initial measure}\la{rem} 
 
  We use {strong uniform} 
Ibragimov-Linnik mixing condition for the simplicity of presentation. 
The {\it uniform} Rosenblatt mixing condition 
\ci{Ros} also is sufficient together with a higher 
degree $>2$  in the bound (\re{med}): there exists $\de >0$  
such that 
$$ 
~~~~~~~~~~~~ 
~~~~~~~~~~~\sup\limits_{x\in\R^n}  
E \Big(\vert v_0(x)\vert^{2+\de}+\vert \nabla u_0(x)\vert^{2+\de} 
 + \vert u_0(x)\vert^{2+\de}\Big) 
<\infty.~~~~~~~~~~~~~~~~~~~~~~~~~~~~~~~~\eqno{(1.8')} 
$$ 
Then (\re{1.12}) requires a modification: 
$$ 
~~~~~~~~~~~~~~~ 
~~~~~~\int _0^\infty r^{n-2}\al^{p}(r)dr <\infty,\,\,\, 
p=\min(\fr\de{2+\de}, \fr 12),~~ 
~~~~~~~~~~~~~~~~~~~~~~~~~~~~~~~~~~~~~~\eqno{(2.12')} 
$$ 
where $\al(r)$ is the  Rosenblatt  
mixing coefficient  defined 
as in  (\re{ilc}) but without $\mu(B)$ in the denominator. 
The statements of Theorem A and their 
proofs remain essentially unchanged, only Lemma \re{il} requires  
a suitable modification \ci{IL}.

\setcounter{equation}{0} 
\section{ Equations with  constant coefficients} 
 
In Sections 3-7 we consider the 
Cauchy problem (\re{1.1}) with  the constant coefficients, 
i.e. 
\beqn 
\left\{\ba{l} 
\ddot u(x,\,t) = \Delta u(x,\,t),~~ x\in \R^n,\\ 
u|_{t=0} = u_0(x),~~ \dot u|_{t=0} = v_0(x). 
\ea 
\right.\la{2.1} 
\eeqn 
Rewrite (\ref{2.1}) in the form 
similar to (\ref{1'}): 
\be\la{2'} 
\dot Y(t)={\cal A}_0 Y(t),\,\,\,t\in\R;\,\,\,\,Y(0)=Y_0. 
\ee 
Here we denote 
\be\la{A0} 
{\cal A}_0=\left( 
 \begin{array}{cc} 
0 & 1\\ 
A_0 & 0 
\end{array}\right), 
\ee 
where 
 $A_0=\De$. 
Denote by $U_0(t),$ $t\in\R,$ the dynamical group for problem 
(\ref{2'}), then $Y(t)=U_0(t)Y_0$. 
Set  $\mu_t(B)=\mu_0(U_0(-t)B)$, 
$B\in {\cal B}({\cal H})$, $t\in\R$. The 
 main result for the problem (\re{2'}) is the following 
\medskip\\ 
{\bf Theorem B}~{\it 
Let $n\ge 3$  be odd,  and Conditions {\bf S0--S3} 
hold. 
Then the conclusions  of Theorem A 
hold with $W =I$, and limiting measure $\mu_\infty$ is  
translation-in\-va\-ri\-ant. 
} 
\medskip 
 
Theorem B can be  deduced  
from Propositions \re{l2.1} and   \re{l2.2} below, 
by using the same arguments as in   
\ci[Thm XII.5.2]{VF}. 
\begin{pro}\la{l2.1} 
The family of  measures $\{\mu_t, t\geq 0\}$ 
is weakly compact in 
 ${\cal H}^{-\ve }$ with any $ \ve >0$, 
and the following  bounds hold: 
\be\la{3.1} 
\sup\limits_{t\ge 0} 
E \Vert U_0(t)Y_0\Vert^2_{R}<\infty,\,\,\,\,\,R>0. 
\ee 
\end{pro} 
\begin{pro}\la{l2.2} 
For any $\Psi\in  {\cal D}$, 
\be\la{2.6} 
\hat \mu_t(\Psi )\equiv\int \exp(i\langle Y,\Psi\rangle )\,\mu_t(dY) 
\rightarrow \exp\{-\fr{1}{2}{\cal Q}_\infty (\Psi ,\Psi)\}, 
~~t\to\infty. 
\ee 
\end{pro}

Propositions \re{l2.1} and  \re{l2.2}  are proved in  
Sections 4 and 5-7, respectively. 
We will use repeatedly the FT formulas (\re{hatA}) and  (\re{tidtx}) 
from Appendix.

\section{Compactness of the measures family} 
\setcounter{equation}{0} 
Here we prove Proposition \re{l2.1} 
with the help of FT.

\subsection{Mixing in terms of the Fourier transform} 
The next proposition reflects the mixing property in
terms of the FT $\hat q^{ij}_0$ of initial CFs $q^{ij}_0$. 
Assumption {\bf S2} implies that  $q^{ij}_0(z)$ 
is a measurable bounded  function. Therefore, it belongs 
to the Schwartz space of tempered distributions as well as 
its FT. 
\begin{pro}\la{l40.1} 
$\hat q^{ij}_0( k)\in L^p(\R^n)$ with $1\le p\le 2$, and 
\be\la{40.7}\int  
(| k|^{2-i-j}+|k|^{-2})|\hat q^{ij}_0( k)|\,d k\le 
C(\varphi)e_0<\infty,\,\,\,i,j=0,1. \ee 
\end{pro} 
{\bf Proof}  We check the bound for  $i=j=1$ 
(in all other cases the proof is similar). By the Bohner Theorem, 
$\hat q^{11}_0d k$ is a nonnegative measure. 
Hence, 
\be\la{40.81}\int |\hat q^{11}_0( k) \,d k| 
= \int \hat q^{11}_0( k)\,d k=q^{11}_0(0)<\infty,\ee 
owing to  {\bf S2}. Similarly, (\ref{1.19}) and (\ref{1.12}) imply 
that 
\beqn\la{40.82} 
\int | k|^{-2}|\hat q^{11}_0( k) \,d k| 
&= &\int | k|^{-2}\hat q^{11}_0( k)\,d k 
= 
C\int |x|^{2-n} q^{11}_0(x)\,dx\nonumber\\ 
&\le& 
C_1 e_0 
\int\limits_0^{+\infty} r\varphi^{1/2}(r) \,dr 
\le C_2\ov\varphi e_0<\infty. 
\eeqn 
It remains to prove that measure $\hat q_0^{11}(k)dk$ 
is absolutely continuous with respect to the Lebesgue measure. 
Function $\phi(r)$ is nonincreasing, hence by (\ref{1.12}) 
\be\la{rpr} 
r^{n-1}\varphi^{1/2}(r)= 
(n-1) 
\int_0^r s^{n-2}\varphi^{1/2}(r)ds\le  
(n-1) 
\int_0^r s^{n-2}\varphi^{1/2}(s)ds\le C\ov\varphi<\infty. 
\ee 
Then (\ref{1.19}) and (\ref{1.12}) imply 
$$ 
\int |q^{11}_0(z)|^2\,dz\le 4e_0^2\int\limits_{\R^n} 
\varphi(|z|)\,dz=  C(n)e_0^2 
 \int_0^\infty r^{n-1}\varphi(r)dr \le C_1 \ov\varphi e_0^2 
 \int _0^\infty \varphi^{1/2}(r)dr <\infty. 
$$ 
Therefore, 
$\hat q^{11}_0( k)\in L^2(\R^n)$. 
\hfill$\Box$ 
\subsection{Proof of the compactness of the family $\{\mu_t\}$} 
We now prove bound (\ref{3.1}).   
Proposition \re{l2.1} then can be deduced  with the help of 
Prokhorov's Theorem \cite[Lemma II.3.1]{VF}, in a way similar to  
\ci[Thm XII.5.2]{VF}. Formulas (\ref{tidtx}),  (\ref{hatA})   and  
Proposition \re{l40.1} imply 
\beqn 
&& 
E\Big(u(x,t) u(y,t)\Big)=:q_t^{00}(x-y) 
\la{4.6kg}\\ 
&&~\nonumber\\ 
&&=\fr 1{(2\pi)^n}\int e^{-i k(x-y) } 
\Bigl[\frac{1+\cos 2|k|  t}{2} 
\hat q_0^{00}( k)+\frac{\sin 2 |k|  t}{2|k| } 
(\hat q_0^{01}( k)+\hat q_0^{10}( k)) 
+\frac{1-\cos 2|k|  t}{2|k| ^2} 
\hat q_0^{11}( k)\Bigr]\,d k,\nonumber\eeqn 
where the integral converges and define a continuous function. 
Similar representations hold for all $i,j=0,1$. 
Therefore,  we have as in (\re{med}), 
\be\la{medt} 
e_t:=q_t^{11}(0) 
-\De q_t^{00}(0)+q_t^{00}(0) 
=\fr 1{(2\pi)^n}\int \Big( 
\hat q_t^{11}( k)+| k|^2\hat q_t^{00}( k)+\hat q_t^{00}( k)\Big)dk. 
\ee 
It remains to  estimate the last integral. 
(\ref{4.6kg}) implies the following representation for  
$\hat q_t^{00}$, 
\be\la{4.6}\hat q_t^{00}(k)=\frac{1+\cos 2|k|  t}{2} 
\hat q_0^{00}( k)+\frac{\sin 2 |k|  t}{2|k| } 
(\hat q_0^{01}( k)+\hat q_0^{10}( k)) 
+\frac{1-\cos 2|k|t}{2|k|^2}\hat q_0^{11}( k).\ee 
Similarly, formulas (\ref{tidtx}), (\ref{hatA}) 
imply 
\be\la{4.611}\hat q_t^{11}(k) 
=|k|^2\frac{1-\cos 2|k|t}{2} 
\hat q_0^{00}(k)-|k|\frac{\sin 2|k|t}{2 } 
(\hat q_0^{01}(k)+\hat q_0^{10}( k)) 
+\frac{1-\cos 2|k|t}{2}\hat q_0^{11}( k).\ee 
Therefore, (\re{40.7}) and (\re{medt}) imply that 
$e_t\le C_1(\varphi)e_0$. Hence, taking expectation in  
(\re{1.5}), we get (\re{3.1}): 
\be\la{3.1b} 
E \Vert U_0(t)Y_0\Vert^2_{R}=  
e_t|B_R|\le 
C_1(\varphi)e_0|B_R|.
\ee 
Here $B_R$ denotes the ball $\{x\in\R^n:~|x|\le R\}$ and  
$|B_R|$ is its volume.   
$\hfill\Box$ 
\begin{cor}\la{QQ} 
Bound (\re{3.1}) implies the convergence of the integrals in 
(\re{qd}). 
\end{cor} 
 
Bound (\re{3.1}) also implies, similarly to (\re{bd}), that 
\be\la{bdt} 
\sup\limits_{t\geq 0}\int\brr Y\brr^2_{\delta}\,\mu_t(dY) 
\leq C_\de(\varphi)e_0  <\infty,~~  \delta >0.\ee 
This integral estimate implies the following corollary
which we will use in Section 9.

\begin{cor}\la{coH} 
{\rm i) Measures $\mu_t$, $t\geq 0$, are concentrated in 
${\cal H}_\de$ for any $\delta>0$, and 
the characteristic functionals $\hat\mu_t$ 
are equicontinuous in 
the dual Hilbert space ${\cal H}_\de^\pr$: 
for all $\Psi_{1},\Psi_2\in {\cal H}_\de^\pr$,
\beqn\la{eqc} 
|\hat\mu_t(\Psi_1)-\hat\mu_t(\Psi_2)| 
&\!\!\!\!\leq&\!\!\!\!\int |\exp(i\langle Y,\Psi_1-\Psi_2\rangle )-1|\mu_t(dY) 
\leq \int |\langle Y,\Psi_1-\Psi_2\rangle |\mu_t(dY)\nonumber\\ 
&\!\!\!\!\leq& \!\!\!\! 
\int \brr Y\brr_\de \cdot\brr \Psi_1-\Psi_2\brr_\de^\pr~\mu_t(dY) 
\leq C(\de,\varphi, e_0)\brr \Psi_1-\Psi_2\brr_\de^\pr,\,\,t\geq 0, 
\eeqn 
where $\brr\cdot\brr_\de^\pr$ denotes the norm in ${\cal H}_\de^\pr$.\\
ii) The quadratic forms 
${\cal Q}_t(\Psi,\Psi)$ are 
 equicontinuous in  ${\cal H}_\de^\pr$:
for all $\Psi_{1},\Psi_2\in {\cal H}_\de^\pr$,
\beqn\la{eqtc} 
\!\!\!\!\!\!\!\!\!\!\!\!\!\!\!\!\!\!&&|{\cal Q}_t(\Psi_1,\Psi_1) -
{\cal Q}_t(\Psi_2,\Psi_2)|\le
C|{\cal Q}_t(\Psi_1-\Psi_2,\Psi_1)+
{\cal Q}_t(\Psi_2,\Psi_1-\Psi_2)|\nonumber\\
\!\!\!\!\!\!\!\!\!\!\!\!\!\!\!\!\!\!&&\le 
C\int |\langle Y,\Psi_1-\Psi_2\rangle|
\Big(|\langle Y,\Psi_1\rangle|+
|\langle Y,\Psi_2\rangle|
 \Big)~
\mu_t(dY)\nonumber\\
\!\!\!\!\!\!\!\!\!\!\!\!\!\!\!\!\!\!&&\le
C\int
\brr Y\brr^2_\de \cdot
\brr \Psi_1-\Psi_2\brr_\de^\pr
\cdot
\Big(
\brr \Psi_1\brr_\de^\pr+
\brr \Psi_2\brr_\de^\pr
 \Big)
~\mu_t(dY)
\le C(\de,\varphi, e_0)\brr \Psi_1-\Psi_2\brr_\de^\pr,\,\,t\geq 0. 
\eeqn
iii) Therefore, the quadratic form 
${\cal Q}_\infty(\Psi,\Psi)$ is 
 continuous in  ${\cal H}_\de^\pr$.
}
\end{cor} 

\subsection{Convergence of the covariance functions} 
Here we prove the convergence 
of the CFs of   measures $\mu_t$. 
This convergence  is used in Section 6.   
\begin{lemma}\la{p40.1} 
The following convergence holds as $t\to\infty$: 
\be\la{40.2} 
q_t^{ij}(z)\to q_{\infty}^{ij}(z),~~\,\,\, 
\forall z\in\R^n,\,\,\,i,j=0,1. 
\ee 
\end{lemma} 
{\bf Proof.} 
(\ref{4.6}) and  (\ref{4.611}) imply  the convergence  
for $i=j$:  the oscillatory terms there 
converge to zero as they 
 are absolutely continuous and summable 
by Proposition \re{l40.1}. 
For $i\ne j$ the proof is similar. 
\hfill$\Box$ 
 
\setcounter{equation}{0} 
\section{Bernstein's argument for the wave equation} 
In this and the subsequent section we develop 
a version of  
Bernstein's `room-corridor' 
method. 
We use the standard integral 
representation  for solutions, 
 divide the domain of integration 
into `rooms' and `corridors' and evaluate  
their contribution. As a result, $\langle  U_0(t)Y_0,\Psi\rangle $  
is represented  
as the sum of 
weakly dependent random variables. 
We  evaluate  the variances  of these  random variables 
which will be important in  next section. 
 
First, we evaluate $\langle  Y(t),\Psi\rangle $ in (\ref{2.6}) 
by using the duality arguments. 
For  $t\in\R$, introduce 
the operators $U'(t)$, $U'_0(t)$ 
in the Hilbert space ${\cal H}'_{\delta}$, 
which are  adjoint to operators 
$U(t)$, $U_0(t)$ in ${\cal H}_{\delta}$. For example, 
\be\la{def} 
\langle Y,U'_0(t)\Psi\rangle = 
\langle U_0(t)Y,\Psi\rangle ,\,\,\, 
\Psi\in {\cal D},   
\,\,\, Y\in {\cal H}_{\delta},\,\,\,t\in\R.   
\ee   
The adjoint groups 
admit a convenient description. 
Lemma \re{ldu} below displays that   
the action of  groups $U'_0(t)$, $U'(t)$ coincides,  
respectively, with the  
action   
of $U_0(t)$, $U(t)$, up to the order of the components.  
In particular, $U'_0(t)$ is a continuous  
group in ${\cal D}$. 
\begin{lemma}\la{ldu} 
For $\Psi=(\Psi^0,\Psi^1)\in {\cal D}$ 
\be\la{UP} 
U'_0(t)\Psi= (\dot\phi(\cdot,t),\phi(\cdot,t)),\,\,\,\,\, 
U'(t)\Psi= (\dot\psi(\cdot,t),\psi(\cdot,t)), 
\ee 
where $\phi(x,t)$ is the solution of Eq.(\ref{2.1}) 
 with the initial datum 
$(u_0,v_0)=(\Psi^1,\Psi^0)$ 
and  
 $\psi(x,t)$ is the solution of Eqn (\ref{1.1}) 
 with the initial date 
$(u_0,v_0)=(\Psi^1,\Psi^0)$. 
\end{lemma} 
{\bf Proof} 
Differentiating (\ref{def}) with 
$Y,\Psi\in {\cal D}$, we obtain 
\be\la{UY} 
\langle  Y,\dot U'_0(t)\Psi\rangle =\langle  \dot U_0(t)Y,\Psi\rangle . 
\ee 
Group $U_0(t)$ has the generator (\ref{A0}) 
The generator of $U'_0(t)$   is the conjugate operator 
\be\la{A0'} 
{\cal A}'_0= 
\left( \begin{array}{cc} 0 & A_0 \\ 
1 & 0 \end{array}\right). 
\ee 
Hence, Eqn (\ref{UP}) holds with 
 $\ddot \psi=A_0\psi$. 
For the group $U'(t)$ the proof is similar. 
 \hfill$\Box$\\ 
 
 Denote $\Phi(\cdot,t)=U'_0(t)\Psi$. Then 
(\ref{def}) means that 
\be\la{defY} 
\langle  Y(t),\Psi\rangle =\langle  Y_0,\Phi(\cdot,t)\rangle, 
\,\,\,\,t\in\R. 
\ee  
{\bf Remark}   The representation (\ref{defY})  plays  
a central role in the proof of Proposition  
\ref{l2.2}. A key observation is that $\Phi(x,t)$ 
is supported by an `inflated' cone of thickness $\approx {\ov r}$ 
where ${\ov r}$ is the diameter of supp~$\Psi$. The last fact follows 
from the {\it strong} Huyghen's principle for  group $U'_0(t)$
which holds for odd $n\ge 3$.
Therefore,  the scalar product $\langle  Y_0,\Phi(\cdot,t)\rangle $  
is represented 
as an integral over the `spherical slab' of width $\approx {\ov r}$. 
This replaces, for a general $n\ge 3$,  
the Kirchhoff integral (\ref{Kir}) written for $n=3$. 
\medskip 
  
Next we introduce `room-corridor'  partition of the space $\R^n$.  
Given $t>0$, choose  
$d\equiv d_t\ge 1$ and 
$\rho\equiv\rho_t>0$. 
Asymptotical relations between $t$, $d_t$ and  $\rho_t$ 
are specified below. 
Define $h=d+\rho$ and 
\be\la{rom} 
a^j=jh,\,\,\,b^j=a^j+d, 
\,\,\,j\in\Z. 
\ee 
We call the slabs $R_t^j=\{x\in\R^n:~a^j\le x^n\le b^j\}$  `rooms' 
and $C_t^j=\{x\in\R^n:~b^j\le x^n\le  a_{j+1}\}$  `corridors'. 
Here  $x=(x^1,\dots,x^n)$, 
$d$ is the width of a room, and 
$\rho$ of a corridor. 
  
Denote by 
$\chi_r$ the indicator of the interval $[0,~ d]$ and 
$\chi_c$ that of $[d,~ h]$ so that 
$\sum_{j\in \Z}(\chi_r(s-jh)+\chi_c(s-jh))=1$ for  
(almost all) $s\in\R$. 
The following  decomposition holds: 
\be\la{res} 
\langle  Y_0,\Phi(\cdot,t)\rangle = \sum_{j\in \Z} 
(\langle  Y_0,\chi_r^j\Phi(\cdot,t)\rangle +\langle  
Y_0,\chi_c^j\Phi(\cdot,t)\rangle ), 
\ee 
where $\chi_r^j:=\chi_r(x^n-jh)$ and  
$\chi_c^j:=\chi_c(x^n-jh)$.  
Consider 
random variables 
 $ r_t^j$, $ c_t^j$, where 
\be\la{100} 
r_t^j= \langle  Y_0,\chi_r^j\Phi(\cdot,t)\rangle  ,~~ 
c_t^j= \langle  Y_0,\chi_c^j\Phi(\cdot,t)\rangle ,~~~~~~j\in\Z. 
\ee 
Then  (\ref{res}) and (\ref{defY}) imply 
\be\la{razli} 
\langle  U_0(t)Y_0,\Psi\rangle =\sum\limits_{j\in\Z} 
(r_t^j+c_t^j). 
\ee 
The series in (\ref{razli})  
is actually a finite sum. 
 In fact,  (\ref{A0'})   and (\ref{Frep}) imply that 
in the Fourier representation,   
$\dot{\hat\Phi}(k,t)=\hat{\cal A}'_0 (k)\hat \Phi(k,t)$ and  
$ 
\hat \Phi(k,t)=\hat{\cal G}'_t( k) 
\hat\Psi(k). 
$ 
Therefore,  
\be\la{frep'} 
\Phi(x,t)=\fr 1{(2\pi)^n} 
\int\limits_{\R^n} e^{-ikx} \hat{\cal G}'_t( k)\hat\Psi( k)~d k. 
\ee 
This can be rewritten as a convolution  
\be\la{conr} 
\Phi(\cdot,t)={\cal R}_t*\Psi, 
\ee  
where ${\cal R}_t=F^{-1}\hat{\cal G}'_t$. 
The support supp$\hspace{0.5mm}\Psi\subset B_{\ov r}$ with an ${\ov r}>0$. 
Then the  convolution representation (\ref{conr}) implies that 
the support of the function $\Phi$ at $t> 0$  
is a subset of an  `inflated light cone' 
\be\la{conp} 
{\rm supp}\hspace{0.5mm}\Phi(x,t)\subset\{(x,t)\in\R^n\times{\R_+}: 
~-{\ov r}\le |x|-t\le{\ov r}\}. 
\ee 
as ${\cal R}_t(x)$ is supported by the light cone  $|x|=t$
as $n$ is odd $\ge \!3$.
The last fact follows from  
the general  Herglotz-Petrovskii formulas  
(see, e.g. \ci[(II.4.4.11) ]{EKS}) 
and is known as {\it strong} Huyghen's principle. 
Finally, (\ref{100})  implies 
\be\la{1000} 
r_t^j= c_t^j= 0             
\,\,\,\quad{\rm     for     }\quad\quad \,jh+t< -{\ov r}\,\,\, 
\quad\mbox{    or    }\quad\,\,jh-t>{\ov r}. 
\ee 
Therefore,  series (\re{razli}) becomes 
a  sum 
\be\la{razl} 
\langle  U_0(t)Y_0,\Psi\rangle =\sum\limits_{-N_t}^{N_t} 
(r_t^j+c_t^j),\,\,\,\,\ds N_t\sim \fr th, 
\ee 
as $h\ge 1$. 
\begin{lemma}  \la{l5.1} 
    Let $n\ge 1$, $m>0$, and {\bf S0--S3} hold. 
The following bounds hold for $t>1$ and ~$\forall j$: 
\be\la{106} 
E|r^j_t|^2\le  C(\Psi)~d_t/t, 
\,\,\,\, 
E|c^j_t|^2\le C(\Psi)~\rho_t/t. 
\ee 
\end{lemma} 
{\bf Proof} 
We discuss the first bound in (\ref{106}) only, the second  
is done in a similar way. 
 
{\it Step 1} 
 Rewrite the left hand side  as the integral of covariance matrices. 
Definition (\ref{100}) and 
Corollary \re{Q}  imply by Fubini's Theorem 
that 
 \be\la{100rq} 
E|r_t^j|^2= \langle  \chi_r^j(x^n)\chi_r^j(y^n)q_0(x-y), 
\Phi(x,t)\otimes\Phi(y,t)\rangle . 
\ee 
The following bound holds true
(cf. \ci[Thm XI.19 (c)]{RS3}): 
\be\la{bphi} 
\sup_{x\in\R^n}|\Phi(x,t)| =\ds{\cal O}(t^{-\fr{n-1}2}), 
\,\,\,\,t\to\infty. 
\ee 
In fact, (\ref{frep'}) and (\ref{hatA}) imply that 
$\Phi$ can be written as the sum 
\be\la{frepe} 
\Phi(x,t)=\fr 1{(2\pi)^n}\sum\limits_{\pm}~~ 
\int\limits_{\R^n} e^{-i(kx\mp|k| t)}  
a^\pm(|k|) 
\hat\Psi( k)~d k, 
\ee 
where $a^\pm(|k|)$  
is a matrix whose entries  
are linear functions in $|k|$  
or $1/|k|$. 
Let us prove the asymptotics (\re{bphi}) along each ray 
$x=vt+x_0$ with   $|v|= 1$, then  it holds uniformly in  
$x\in\R^n$ owing to (\ref{conp}).  
In polar coordinates, we get from (\ref{frepe}), 
\be\la{freper} 
\Phi(vt+x_0,t)=\fr 1{(2\pi)^n}\sum\limits_{\pm}~~ 
\int_0^\infty \left( 
\int\limits_{|k|=r} e^{-i(kv\mp|k|)t-ikx_0}  
a^\pm(|k|) 
\hat\Psi( k)~d S(k)\right)dr. 
\ee 
This is a sum of oscillatory integrals with the phase 
functions $\phi_\pm(k)=kv\pm |k|$.  
The standard form of the method of stationary  phase 
is not applicable here 
as the set of stationary points $\{k\in\R^n:~ 
\na\phi_\pm(k)=0\}$, is a ray $v=\pm k/|k|$, and 
the Hessian  is degenerate everywhere. 
On the other hand, restricted to the sphere $|k|=r$ 
with a fixed $r>0$, each phase function  
$\phi_\pm^r:=\left.\phi_\pm\right|_{|k|=r}=kv\pm r$, 
has two  
stationary points $\pm vr$, and the Hessian is nondegenerate 
everywhere: 
\be\la{Hess} 
  {\rm rank}\left({\rm Hess}~\phi_\pm^r(k)\right)=n-1,\,\,\,\,|k|=r. 
\ee 
Hence, the inner integral in  (\ref{freper}) 
is ${\cal O}(t^{-(n-1)/2})$ 
according  to the standard method of 
stationary phase, \ci{F}. 
At last, 
for the integral in $r$ in  (\ref{freper}), has the same  
asymptotics in $t$  as  $\hat\Psi(k)$ decay rapidly 
at infinity.  
 
{\it Step 2}  
 According to 
(\ref{conp})  and 
(\ref{bphi}), Eqn (\ref{100rq}) 
implies that 12
\be\la{er} 
E|r_t^j|^2\le C t^{-n+1}\int 
\limits_{S_t^{\ov r}\times S_t^{\ov r}} 
\chi_r^j(x^n) 
\Vert q_0(x-y)\Vert ~ dxdy, 
\ee 
where $S_t^{\ov r}$ is an `inflated sphere' 
$\{x\in\R^n:~-{\ov r}\le |x|-t\le {\ov r}\}$ 
and $\Vert q_0(x-y)\Vert$ stands for the norm of  
the $2\times 2$-matrix, $\left(q_0^{ij}(x-y)\right)$. 
The  estimate  (\ref{1.19}) implies then 
\be\la{erp} 
E|r_t^j|^2\le C t^{-n+1}\int 
\limits_{S_t^{\ov r}\times S_t^{\ov r}} 
\chi_r^j(x^n) 
 \varphi^{1/2}(|x-y|) 
~ dxdy, 
\ee 
For large  $t$, 
this integral can be reduced to the product of the 
spheres $S_t=\{x\in\R^n:~|x|=t\}$: 
\be\la{ers} 
E|r_t^j|^2\le C_1{\ov r}^2 t^{-n+1} 
\int\limits_{S_t}\chi_r^j(x^n) 
\left(\int\limits_{S_t} \varphi^{1/2}(|x-y|) 
~ dS(y)\right)dS(x), 
\ee 
where $dS$ is a Lebesgue measure on the sphere. 
The inner integral  can be estimated  
by a direct computation owing to (\ref{1.19}): 
\be\la{qphi} 
\int\limits_{S_t} 
 \varphi^{1/2}(|x-y|)~ dS(y) 
= 
C(n)\int_0^{2t} r^{n-1} 
 \varphi^{1/2}(r) dr\le C(n)\ov\varphi,~~~~ 
\,\,\,\,t\ge 0. 
\ee 
Therefore, (\ref{ers}) and (\ref{1.12}) imply 
$$ 
~~~~~~~~~~~~~~~~~~~~~~~~~~~~~~~~~~E|r_t^j|^2\le C_1(n){\ov r}^2 t^{-n+1} 
\int\limits_{S_t}\chi_r^j(x^n) 
dS(x)\le 
 C_2(n)~d_t/t.~~~~~~~~~~~~~~~~~~~~~ 
~~~~~~~~ 
\Box 
$$ 
 

\setcounter{equation}{0} 
\section{Convergence of the characteristic functionals} 
In this section we complete the proof of Proposition \ref{l2.2}. 
As was said, we use a version of the CLT 
developed by Ibragimov and Linnik. This gives the convergence 
to an equilibrium Gaussian measure. 
If ${\cal Q}_{\infty}(\Psi,\Psi)=0$, Proposition \ref{l2.2} 
is obvious. Thus, we may assume that 
\be\la{5.*} 
{\cal Q}_{\infty}(\Psi,\Psi)\not=0. 
\ee 
Choose  $0<\de<1$ and
\be\la{rN} 
\rho_t\sim t^{1-\delta}, 
~~~d_t\sim\fr t{\ln t},~~~~\,\,\,t\to\infty. 
\ee 
\begin{lemma}\la{r} 
The following limit holds true: 
\be\la{7.15'} 
N_t\Bigl( 
\varphi(\rho_t)+\Bigl( 
\frac{\rho_t}{t}\Bigr)^{1/2}\Bigr) 
+ 
N_t^2\Bigl( 
\varphi^{1/2}(\rho_t)+\frac{\rho_t}{t}\Bigr) 
\to 0 ,\quad t\to\infty. 
\ee 
\end{lemma} 
{\bf Proof}. 
Eqn (\ref{7.15'}) follows from (\ref{rpr}) as 
(\ref{rN}) and (\ref{razl}) imply that $N_t\sim\ln t$. 
\hfill$\Box$\\ 
 
By the triangle inequality, 
\beqn 
|\hat\mu_t(\Psi) -\hat\mu_{\infty}(\Psi)|&\le& 
|E\exp\{i \langle  U_0(t)Y_0,\Psi\rangle  \}- 
E\exp\{i{{\sum}}_t r_t^j\}| 
\nonumber\\ 
&&+|\exp\{-\frac{1}{2}{\sum}_t E|r_t^j|^2\} - 
\exp\{-\frac{1}{2} {\cal Q}_{\infty}(\Psi,\Psi)\}| 
\nonumber\\ 
&& 
+ |E \exp\{i{\sum}_t r_t^j\} - 
\exp\{-\frac{1}{2}{\sum}_t E|r_t^j|^2\}|\nonumber\\ 
&\equiv& I_1+I_2+I_3, \la{4.99} 
\eeqn 
where the sum ${\sum}_t$ stands for $\sum\limits_{j=-N_t}^{N_t}$. 
We are going to   show  that all summands  
$I_1$, $I_2$, $I_3$  
 tend to zero 
 as  $t\to\infty$.\\ 
{\it Step (i)} 
Eqn (\ref{razl}) implies 
\be\la{101} 
 I_1=|E\exp\{i{\sum}_t r^j_t \} 
\ds{(\exp\{i{\sum}_t c^j_t\}-1)|\le 
 {\sum}_t E|c^j_t|\le 
{\sum}_t(E|c^j_t|^2)^{1/2}}. 
\ee 
From (\ref{101}), (\ref{106})  and (\ref{7.15'}) we obtain that 
\be\la{103} 
I_1\le C N_t(\rho_t/t)^{1/2}\to 0,~~t\to \infty. 
\ee 
{\it Step (ii)} 
By the triangle inequality, 
\beqn 
I_2&\le& \frac{1}{2} 
|{\sum}_t E|r_t^j|^2- 
 {\cal Q}_{\infty}(\Psi,\Psi) | 
\le 
 \frac{1}{2}\, 
|{\cal Q}_{t}(\Psi,\Psi)-{\cal Q}_{\infty}(\Psi,\Psi)| 
\nonumber\\ 
&&+ \frac{1}{2}\, |E\Bigl({\sum}_t r_t^j\Bigr)^2 
-{\sum}_tE|r_t^j|^2| + 
 \frac{1}{2}\, |E\Bigl({\sum}_t r_t^j\Bigr)^2 
-{\cal Q}_{t}(\Psi,\Psi)|\nonumber\\ 
&\equiv& I_{21} +I_{22}+I_{23}\la{104}, 
\eeqn 
where ${\cal Q}_{t}$ is  a quadratic form with  
the integral kernel $\Big(Q_t^{ij}(x,y)\Big)$. 
Eqn (\ref{40.2}) implies that 
 $I_{21}\to 0$. 
As to  $I_{22}$, we first  have that 
\be\la{i22} 
I_{22}\le \sum\limits_{j< l} 
 E|r_t^j r_t^l|. 
\ee 
The next lemma   is a corollary of \ci[Lemma 17.2.3]{IL}. 
\begin {lemma}\la{il} 
 Let $ \xi$ be a complex random value  
 measurable with respect to 
$\sigma$-algebra $\sigma({\cal A})$, 
  $\eta$  with respect to 
$\sigma$-algebra $\sigma({\cal B})$, 
and {\rm dist}$({\cal A}, {\cal B})\ge r>0$. \\  
i) Let $(E|\xi|^2)^{1/2}\le a$, $(E|\eta|^2)^{1/2}\le b$. Then 
$$ 
|E\xi\eta-E\xi E\eta|\le 
C ab~ 
\varphi^{1/2}(r). 
$$ 
ii) Let $|\xi|\le a$, $|\eta|\le b$ almost sure. Then 
$$ 
|E\xi\eta-E\xi E\eta|\le Cab~ 
\varphi(r). 
$$ 
\end{lemma} 
\bigskip 
 
We apply Lemma \re{il} to deduce that 
$I_{22}\to 0$ as $t\to\infty$. 
Note  that 
$r_t^j= 
\langle  Y_0(x),\chi_r^j(x^n)({\cal R}_t*\Psi)\rangle $ is measurable 
with respect to the $\sigma$-algebra  $\sigma(R_t^j)$. 
The distance 
between the different rooms $R_t^j$ 
is greater or equal to 
$\rho_t$ according to  
 (\ref{rom}). 
Then (\ref{i22}) and {\bf S1}, {\bf S3} imply,  
together with  
Lemma \ref{il} i), that 
\be\la{i222} 
I_{22}\le 
C N_t^2\varphi^{1/2}(\rho_t), 
\ee 
which goes to $0$ as $t\to\infty$ because of  
 (\ref{106}) and (\ref{7.15'}). 
Finally,  it remains to check  
that $I_{23}\to 0$, 
$t\to\infty$. By the 
Cauchy - Schwartz inequality, 
\beqn 
I_{23}&\le& 
 |E\Bigl({\sum}_t r_t^j\Bigr)^2 
- E\Bigl({\sum}_t r_t^j + 
{\sum}_t c_t^j\Bigr)^2 |\nonumber\\ 
& \le& 
C N_t{\sum}_t E |c_t^j|^2  + 
C\Bigl( 
E({\sum}_t r_t^j)^2\Bigr)^{1/2} 
\Bigl( 
N_t{\sum}_t E|c_t^j|^2\Bigr)^{1/2}.\la{107} 
\eeqn 
Then  (\ref{106}), (\ref{i22}) and (\ref{i222}) 
imply 
$$ 
E({\sum}_t r_t^j)^2\le 
{\sum}_tE|r_t^j|^2 +2{\sum}_{j<l}E|r_t^j r_t^l| 
\le 
CN_td_t/t+C_1N_t\varphi^{1/2}(\rho_t)\le C_2<\infty. 
$$ 
Now 
(\ref{106}),  (\ref{107}) and (\ref{7.15'}) yields 
\be\la{106'} 
I_{23}\le C_1  N_t^2\rho_t/t+C_2 N_t(\rho_t/t)^{1/2} \to 0,~~t\to \infty. 
\ee 
So,  all terms $I_{21}$, $I_{22}$, $I_{23}$  
in  (\ref{104}) 
tend to zero. 
Then 
(\ref{104}) implies that 
\be\la{108} 
I_2\le 
\frac{1}{2}\, 
|{\sum}_{t}E|r_t^j|^2- 
 {\cal Q}_{\infty}(\Psi,\Psi)| 
\to 0,~~t\to\infty. 
\ee 
{\it Step (iii)} 
It remains to verify that 
\be\la{110} 
I_3=| 
E\exp\{i{\sum}_t r_t^j\} 
-\exp\{-\fr12E\Big({\sum}_t r_t^j\Big)^2\}| \to 0,~~t\to\infty. 
\ee 
Using Lemma \ref{il}, ii) 
we obtain: 
\beqn 
&&|E\exp\{i{\sum}_t r_t^j\}-\prod\limits_{-N_t}^{N_t} 
E\exp\{i r_t^j\}| 
\nonumber\\ 
&\le& 
|E\exp\{ir_t^{-N_t}\}\exp\{i\sum\limits_{-N_t+1}^{N_t} r_t^j\}  - 
 E\exp\{ir_t^{-N_t}\}E\exp\{i\sum\limits_{-N_t+1}^{N_t} r_t^j\} | 
\nonumber\\ 
&&+ 
|E\exp\{ir_t^{-N_t}\}E\exp\{i\sum\limits_{-N_t+1}^{N_t} r_t^j\} 
-\prod\limits_{-N_t}^{N_t} 
E\exp\{i r_t^j\}| 
\nonumber\\ 
&\le& C\varphi(\rho_t)+ 
|E\exp\{i\sum\limits_{-N_t+1}^{N_t} r_t^j\} 
-\prod\limits_{-N_t+1}^{N_t} 
E\exp\{i r_t^j\}|.\nonumber 
\eeqn 
We then apply Lemma \ref{il}, ii) recursively 
and get, according to  Lemma \ref{r}, 
\be\la{7.24'} 
|E\exp\{i{\sum}_{t} r_t^j\}-\prod\limits_{-N_t}^{N_t} 
E\exp\{i r_t^j\}| 
\le 
C N_t\varphi(\rho_t)\to 0,\quad t\to\infty. 
\ee 
It remains to check that 
\be\la{110'} 
|\prod\limits_{-N_t}^{N_t} E\exp\{ir_t^j\} 
-\exp\{-\fr12{\sum}_{t} E|r_t^j|^2\}| \to 0,~~t\to\infty. 
\ee 
According to the standard statement of the CLT 
(see, e.g. \ci[Thm 4.7]{P}), 
it suffices to verify the Lindeberg condition: 
$\forall\ve>0$ 
\be\la{Lind} 
\frac{1}{\sigma_t} 
{\sum}_t 
 E_{\ve\sqrt{\sigma_t}} 
|r_t^j|^2 \to 0,~~t\to\infty. 
\ee 
Here $\sigma_t\equiv {\sum}_t E |r^j_t|^2,$ 
and $E_\de f\equiv E X_\de f$, where $X_\de$ is 
the indicator of the event $|f|>\de^2.$ 
Note that (\ref{108}) and (\re{5.*}) imply that 
$$\sigma_t \to 
{\cal Q}_{\infty}(\Psi,\Psi)\not= 0,~~t\to\infty.$$ 
Hence it remains to verify that $\forall\ve>0$ 
\be\la{linc} 
{\sum}_t E_{\ve} 
|r_t^j|^2 \to 0,~~t\to\infty. \ee 
We check (\ref{linc}) in Section 7. This will complete 
the proof of Proposition \ref{l2.2}. 
\hfill$\Box$

 
\setcounter{equation}{0} 
\section{The Lindeberg condition} 
The proof of (\ref{linc}) can be reduced to the case when 
for some $\La\ge 0$ we have, almost sure that  
 \be\la{ass1} 
  |u_0(x)|+|v_0(x)|\leq \La<\infty, ~~~x\in\R^n. 
  \ee 
Then the proof of (\ref{linc}) is reduced  to 
the convergence 
\be\la{111} 
{\sum}_t E|r_t^j|^4 \to 0,~~t\to\infty 
\ee 
by using Chebyshev's inequality. 
The general case can be covered by standard cutoff arguments 
taking into account that   
bound  (\ref{106}) for $E|r^j_t|^2$ 
depends only on $e_0$ and $\varphi$. 
 The last fact is evident  from  
(\ref{er})-(\ref{qphi}). 
We deduce (\ref{111}) from  
\begin{theorem}  \la{p5.1} 
Let the conditions of Theorem B  hold  
and assume that (\ref{ass1}) is fulfilled. 
Then 
for any $\Psi\in {\cal D}$ 
there exists a constant $C(\Psi)$ 
such that 
\be\la{112} 
E|r_t^j|^4\le 
C(\Psi) \La^4d_t^2/t^2,   ~~t>1. 
\ee 
\end{theorem} 
{\bf Proof.} {\it Step 1}~ 
Given four points $x_1,x_2,x_3,x_4\in\R^n$, set: 
$$
M_0^{(4)}(x_1,...,x_4)= 
E\left(Y_0(x_1)\otimes... \otimes Y_0(x_4)\right).
$$ 
Then, similarly to (\ref{100rq}), 
Eqns (\ref{ass1}) and (\ref{100}) imply by 
the Fubini Theorem that  
\be\la{11.2} 
E|r_t^j|^4= 
\langle  \chi_r^j(x_1^n) \ldots \chi_r^j(x_4^n)M_0^{(4)}(x_1,\dots,x_4), 
\Phi(x_1,t)\otimes\dots\otimes\Phi(x_4,t)\rangle. 
\ee 
Let us analyse the domain of the integration ${(\R^n)^4}$ in 
the RHS of (\ref{11.2}). 
We partition  ${(\R^n)^4}$ into three parts $W_2$,  $W_3$  and $W_4$: 
\be\la{Wi} 
{(\R^n)^4}= 
\bigcup\limits_{i=2}^4 W_i,\quad 
W_i=\{\bar x=(x_1,x_2,x_3,x_4)\in {(\R^n)^4}: 
|x_1-x_i|=\max\limits_{p=2,3,4}|x_1-x_p|\}. 
\ee 
Furthermore,  given  $\bar x=(x_1,x_2,x_3,x_4)\in W_i$, 
 divide $\R^n$ into three parts 
$S_j$ , $j=1,2,3$: $\R^n=S_1\cup S_2\cup S_3$,   
by two 
hyperplanes orthogonal to 
the segment $[x_1,x_i]$ and  partitioning  it into three equal segments, 
where  $x_1\in S_1$ and $x_i\in S_3$. 
Denote by 
$x_p$, $x_q$  the two remaining points  
with $p,q\ne 1,i$. 
Set: 
 ${\cal A}_i=\{\bar x\in W_i:~ 
 x_p\in S_1,  x_q\in S_3 \}$, 
 ${\cal B}_i=\{\bar x\in W_i:~ 
 x_p, x_q\not\in S_1 \}$ and 
 ${\cal C}_i=\{\bar x\in W_i:~ 
 x_p, x_q\not\in S_3 \}$, $i=2,3,4$. 
 Then 
$W_i={\cal A}_i\cup {\cal B}_i\cup {\cal C}_i$. 
Define the function ${\rm m}^{(4)}_0(\bar x)$, $\bar x\in{(\R^n)^4},$ 
in the following way: 
\beqn\la{M} 
{\rm m}^{(4)}_0(\bar x)\Bigr|_{W_i}= 
\left\{ 
\ba{ll} 
M_0^{(4)}(\bar x)- 
q_0(x_1-x_p)\otimes 
 q_0(x_i-x_q),\quad \bar x\in{\cal A}_i,\\ 
M_0^{(4)}(\bar x),\quad \bar x\in{\cal B}_i\cup {\cal C}_i. 
\ea 
\right. 
\eeqn 
This determines ${\rm m}^{(4)}_0(\bar x)$ 
correctly   for almost all quadruples $\bar x$. 
Note that 
\beqn 
\!&&\!\!\!\langle 
  \chi_r^j(x_1^n) \ldots \chi_r^j(x_4^n) 
q_0(x_1-x_p)\otimes q_0(x_i-x_q), 
\Phi(x_1,t)\otimes\dots\otimes\Phi(x_4,t)\rangle  
\nonumber\\ 
\!&&\!\!\!\!\!\!\!\!\!=\langle  
 \chi_r^j(x_1^n)\chi_r^j(x_p^n) q_0(x_1-x_p), 
\Phi(x_1,t)\otimes\Phi(x_p,t)\rangle~  
\langle  \chi_r^j(x_i^n)\chi_r^j(x_q^n) q_0(x_i-x_q), 
\Phi(x_i,t)\otimes\Phi(x_q,t)\rangle . 
\nonumber 
\eeqn 
Each factor here is bounded by 
$ C(\Psi)~d_t/t$. 
Similarly to  (\ref{106}), 
this can be deduced  from an expression of type (\ref{100rq}) 
for the factors. 
Therefore, the proof of (\ref{112}) reduces to the proof 
of the  bound 
\be\la{st1} 
I_t:=|\langle  \chi_r^j(x_1^n)\ldots\chi_r^j(x_4^n){\rm 
m}^{(4)}_0(x_1,\dots,x_4), 
\Phi(x_1,t)\otimes\dots\otimes\Phi(x_4,t)\rangle |\le 
C(\Psi) \La^4d_t^2/t^2,\quad t > 1. 
\ee 
{\it Step 2}~ 
Similarly to (\re{er}), 
 Eqn (\ref{bphi}) implies,  
\be\la{500} 
I_t\le 
 C(\Psi)~t^{-2n+2}\int\limits_{ (S_t^{\ov r})^4 } 
\chi_r^j(x_1^n) \ldots \chi_r^j(x_4^n) 
 |{\rm m}^{(4)}_0(x_1,\dots,x_4)| dx_1~dx_2~dx_3~ dx_4, 
\ee 
where $S_t^{\ov r}$ is an `inflated sphere' 
$\{x\in\R^n:~-{\ov r}\le |x|-t\le {\ov r}\}$. 
Let us estimate ${\rm m}^{(4)}_0$ 
using Lemma  \ref{il}, ii). 
\begin{lemma}\la{l11.1} 
For each $i=2,3,4$ and almost all  $\ov x\in W_i$ the  
following bound holds 
\be\la{115} 
 |{\rm m}^{(4)}_0(x_1,\dots,x_4)|  
\le 
C \La^4\varphi(|x_1-x_i|/3). 
\ee 
\end{lemma} 
{\bf Proof.} 
For $\bar x\in {\cal A}_i$ we apply  
 Lemma  \ref{il}, ii)  
to $\R^2\otimes\R^2$-valued random variables  
$\xi=Y_0(x_1)\otimes Y_0(x_p)$ and  
 $\eta=Y_0(x_i)\otimes Y_0(x_q)$. Then  
(\ref{ass1}) implies the bound  for almost all $\bar x\in {\cal A}_i$ 
\be\la{M1} 
|{\rm m}^{(4)}_0(\bar x)|\le 
C \La^4 \varphi(|x_1-x_i|/3). 
\ee 
For  $\bar x\in {\cal B}_i$, 
we apply  
 Lemma  \ref{il}, ii)  
to  
$\xi=Y_0(x_1)$ and  
 $\eta= Y_0(x_{p})\otimes Y_0(x_{q})\otimes Y_0(x_{i})$. Then  
  {\bf S0}  implies a similar bound for  
almost all $\bar x\in {\cal B}_i$, 
\be\la{113} 
|{\rm m}^{(4)}_0(\bar x)|= 
| M_0^{(4)}(\bar x) -  
EY_0(x_1) \otimes 
E\Bigl( Y_0(x_{p})\otimes Y_0(x_{q})\otimes Y_0(x_{i})\Bigr)| 
\le C \La^4 \varphi(|x_1-x_i|/3), 
\ee 
and the same for  
almost all  
$\bar x\in {\cal C}_i$. \hfill$\Box$ 
\medskip\\ 
{\it Step 3}~It remains to prove the following bounds  
for each $i=2,3,4$ (cf (\re{erp})): 
\be\la{vit} 
V_i(t):=\int\limits_{ (S_t^{\ov r})^4 } 
\chi_r^j(x_1^n) \ldots \chi_r^j(x_4^n) 
 X_i(\ov x)   \varphi(|x_1-x_i|/3)  
dx_1~dx_2~dx_3~ dx_4  
\le Cd_t^2 t^{2n-4}, 
\ee 
where $X_i$ is an indicator of the set $W_i$. 
In fact, this integral does not depend on $i$,  hence 
set $i=2$ in the integrand. 
Similarly to (\re{ers}), 
this integral  can be reduced to the product of four spheres 
$S_t=S_t^0$:  for large $t$, 
\be\la{500s} 
V_i(t) 
\le 
 C{\ov r}^4\int\limits_{ (S_t)^2 } 
\chi_r^j(x_1^n) \varphi(|x_1-x_2|/3)  
\left[\int\limits_{ S_t }\chi_r^j(x_3^n)  
\left(\int\limits_{ S_t } 
X_2(\ov x)~dS(x_4)\right)~dS(x_3) \right]    
 dS(x_1)~dS(x_2). 
\ee 
Now a key observation is that the inner integral  in $dS(x_4)$ 
is ${\cal O}(|x_1-x_2|^{n-1})$ as $X_2(\ov x)=0$ for  
$|x_4-x_1| > |x_1-x_2|$. This implies 
\be\la{500s4} 
V_i(t) 
\le 
 C{\ov r}^4\int\limits_{ S_t }\chi_r^j(x_1^n) 
\left(\int\limits_{ S_t } 
 \varphi(|x_1-x_2|/3) 
|x_1-x_2|^{n-1}~ dS(x_2)\right)~dS(x_1) 
 \int\limits_{ S_t }\chi_r^j(x_3^n)  
~dS(x_3). 
\ee 
The inner integral in $dS(x_2)$ can be estimated  
by a direct computation: similarly to  
 (\re{qphi}),  
\beqn\la{qphi4} 
&&\int\limits_{ S_t } 
 \varphi(|x_1-x_2|/3) 
|x_1-x_2|^{n-1}~ dS(x_2) 
=C(n)\int_0^{2t}r^{2n-3} \varphi( r/3)~dr\nonumber\\ 
&&\le 
C_1(n) \sup\limits_{r\in [0,2t]} ~r^{n-1}\varphi^{1/2}( r/3) 
\int_0^{2t}r^{n-2} \varphi^{1/2}( r/3)~dr. 
\eeqn 
The `$\sup$' and the last integral are bounded by  
(\re{rpr}) and (\re{1.12}), respectively. Therefore,  (\re{vit}) 
follows from (\re{500s4}). 
This completes the proof of 
Theorem \ref{p5.1}. 
\hfill$\Box$\\ 
~\\ 
{\bf Proof of convergence (\ref{111}).} 
The estimate (\ref{112}) implies, since  $d_t\le h\sim t/N_t$, 
$$ 
~~~~~~~~~~~~~~~~~~~~~~~~~~~~~~{\sum}_t 
E|r_t^j|^4  \le 
\fr{C\La^4d_t^2} 
{t^2}N_t 
\le 
\fr{C_1\La^4} 
{N_t} 
\to 0,\quad N_t\to\infty.~~~~~~~~~~~~~~ 
~~~~~~~~~~~~~~~~~~~~~~~~~~~\Box 
$$

\setcounter{equation}{0} 
\section{Scattering theory for infinite energy solutions} 
 
As was said in Sections 1-2, 
  we reduce the proof of Theorem A to Theorem B 
by using a  special 
version of the scattering theory, 
for solutions of infinite energy. 
Recall, 
$U(t)$, $U_0(t)$ 
are the dynamical groups of 
the Cauchy problems (\ref{1'}), 
(\ref{2'}), respectively. 
\begin{theorem} \la{t6.1} 
Let {\bf E1}-{\bf E3} hold, and  $n\geq 3$ be odd. Then 
there exist $\delta, \gamma>0$ and linear continuous 
operators 
$\Theta, \rho(t):~{\cal H}_{\delta}\to {\cal H}$ 
such that for $Y_0\in {\cal H}_{\delta}$ 
\be\la{6.2} 
U(t) Y_0=\Theta  U_0(t) Y_0+\rho(t)Y_0,\,\,\,t\geq 0, 
\ee 
and for any $R>0$ there exists a constant 
$C=C(R,\delta,\gamma)$,  such that for $Y_0\in{\cal H}_\de$ 
\be\la{6.2'} 
\Vert \rho(t)Y_0\Vert_{R} 
\le C~ e^{-\gamma t}\brr Y_0\brr_{\delta},\,\,\,t\geq 0. 
\ee 
\end{theorem} 
 
We deduce Theorem \re{t6.1} 
with the help of 
 duality from a special version of the finite energy scattering 
theory that is developed below. 
Denote $\brr\cdot\brr'_\de$ the norm in the Hilbert space 
${\cal H}_\de'$, dual to ${\cal H}_\de$. 
\begin{lemma}\la{p6.1}   
The following bound holds true:   
\be\la{6.5}   
\brr U'_0(t)\Psi\brr'_{\delta}\le   
C e^{\de |t|}\brr\Psi\brr'_{\delta},~~   
\forall \Psi\in {\cal H}'_{\delta},~~t\in\R.   
\ee   
\end{lemma}   
   
Lemma \re{p6.1} follows by duality from Lemma \re{l6.1}:   
\begin{lemma}\la{l6.1}   
Let  {\bf E1}--{\bf E2} hold. Then $\forall \delta> 0$   
the operator $U_0(t)$ is continuous in ${\cal H}_{\delta}$, and   
there  exists a constant  $C=C(\delta)>0$  such that   
for $Y_0\in {\cal H}_{\delta}$   
\be\la{6.3}   
\brr U_0(t) Y_0\brr_{\delta}\le   
Ce^{\de |t|}\brr Y_0\brr_{\delta},\,\,\,t\in\R.   
\ee   
\end{lemma}   
{\bf Proof.}   
It suffices to consider  $Y_0\in {\cal D}$.   
Then $U_0(t) Y_0\in {\cal D}$.   
Denote   
$$   
E_{\delta}(Y)\equiv   
\int e^{-2\de |x|}   
\Big(\vert v(x)\vert ^2+|\nabla u(x)|^2\Big)\, 
dx,\,\,\,Y=(u,v)\in {\cal H}_{\delta}.   
$$   
Then   the derivative 
$$   
\dot E_{\delta}(U_0(t) Y_0)= 2   
\int e^{-2\de |x|}   
\Big( \dot u(x,t)\ddot u(x,t) + \nabla u(x,t) \nabla\dot u(x,t)  
\Big)\,dx.   
$$   
Substituting $\ddot u(x,t)=\triangle u(x,t)$   
and integrating by parts, we obtain   
$$   
\dot E_{\delta}(U_0(t) Y_0)=-2   
\int   
\nabla e^{-2\de |x|}   
\cdot  \nabla u(x,t)\dot u(x,t) \,dx.   
$$   
Then   
$|\dot E_{\delta}(U_0(t) Y_0)|\le 2\de E_{\delta}(U_0(t) Y_0)$   
by  the Young inequality.   
Therefore, the   
Gronwall inequality implies   
$$   
E_{\delta}(U_0(t) Y_0)\le   
e^{2\de |t|}   E_\de( Y_0).   
$$   
In other words,   
\be\la{6.4}   
\int e^{-2\de |x|}   
\Big(\vert \dot u(x,t)\vert ^2 +\vert\nabla u(x,t)\vert ^2 \Big)\,dx   
\le   e^{2\de |t|}   E_{\delta}( Y_0).   
\ee   
It remains to estimate  the norm $\br u(\cdot,t)\br_{\delta}$, 
where 
$$   
\br u(\cdot,t)\br_{\delta}^2\equiv\int \exp{(-2\de |x|)}   
|u(x,t)|^2\,dx.   
$$   
In fact, we have:   
$$   
\br u(\cdot,t)\br_{\delta}\le   
\br u^0(x)\br_{\delta}+   
\br\int\limits_0^t \dot u(\cdot,{\ov r})\, d{\ov r}\br_{\delta}\le   
\br u^0(x)\br_{\delta}+   
|\int\limits_0^t \br\dot u(\cdot,{\ov r})\br_{\delta}\, d{\ov r} |.   
$$   
Using (\ref{6.4}),  we get   
$$   
~~~~~~~~~~~~~~~~~~~~~~~ 
~~~~~~~~~~~~~~~~~~~~\br u(\cdot,t)\br_{\delta}\le   
Ce^{\delta |t|}\brr Y_0\brr_{\delta}. ~~~~~~ 
~~~~~~~~~~~~~~~~~~~~~~~~~~~~~~~~~~~~~~~~~ 
 \Box    
$$

Now we employ  Vainberg's bounds for  the local  
energy decay;   
this plays the key role in the proof of Theorem  \re{t6.1}.   
We use the Sobolev space  
${\cal H}_{(R)}=H^1(B_R)\oplus L^2(B_R)$   
with the norm $\Vert\cdot\Vert_{(R)}$, $R>0$. 
Recall that  $\stackrel{0\,\,\,\,\,\,\,} 
{H^{-1}}\!\!(B_R)$ is a completion of   
$D_R=\{\psi\in D: \supp\psi\subset B_R\}$ in the Hilbert   
norm of the   
Sobolev space $H^{-1}(\R^n)$.  
We will use the following convenient description 
of a dual space  (see, e.g. \ci[Scn I.3.4]{VP}).   
\begin{lemma}   
$\stackrel{0\,\,\,\,\,\,\,}{H^{-1}}\!\!(B_R)$ is the dual  
to the Hilbert space  
$H^1(B_R)$   
with respect to the scalar product $\langle \cdot,\cdot\rangle $.   
\end{lemma}   
\begin{cor}   
The dual space to the Hilbert space ${\cal H}_{(R)}$   
with respect to  $\langle \cdot,\cdot\rangle $ is   
\be\la{dual}   
{\cal H}_{(R)}^\pr=\stackrel{0\,\,\,\,\,\,\,}{H^{-1}}\!\!(B_R)\oplus 
 L^2(B_R).   
\ee   
\end{cor}   

Note that ${\cal H}_{(R)}^\pr$ is a subspace of  
${\cal H}'_\de$ with any $\de\in\R$.

\begin{definition}   
 ${\cal H}'$ denotes the space $\cup_{R>0}{\cal H}_{(R)}^\pr$ 
endowed with the following convergence:   
a  sequence $\Psi_n$ converges to $\Psi$ in ${\cal H}'$ iff   
$\exists R>0$ s.t. all  $\Psi_n\in {\cal H}'_{(R)}$, and   
$\Psi_n$ converge to $\Psi$ in the norm of ${\cal H}'_{(R)}$.   
\end{definition}   
 
 Below, we    
consider   
 the continuity of the maps from   
${\cal H}'$  only in the sense of the sequential   
continuity.   
Vainberg's results  imply the following   
lemma which we  prove in Appendix.   
\begin{lemma} \la{l6.2}   
Let {\bf E1}--{\bf E3} hold, and let $n\ge 3$ be odd.   
 Then   
$\forall R,R_0>0$  there exist constants $\al, C(R,R_0)>0$  
and  
 $ T=T(R,R_0)>0$  
 such that for  $\Psi\in {\cal H}_{(R)}'$   
\be\la{6.6}   
\Vert U'(t) \Psi\Vert_{L^2(B_{R_0})\oplus H^1(B_{R_0})}\le   
C(R,R_0) e^{-\al t}   
\Vert \Psi\Vert'_{(R)},~~t\ge T,   
\ee   
where $\Vert\cdot\Vert'_{(R)}$ denotes the norm of the dual  
Hilbert space    
${\cal H}_{(R)}^\pr$.   
\end{lemma}   
   
Now  we are in position to discuss our version of the     
 scattering theory for  solutions of finite   
energy.   
   
\begin{pro}\la{t6.2}   
Let {\bf E1}--{\bf E3} hold, and $n\ge 3$ be odd.   
Then there exist   
$\delta, \gamma>0$ and   
linear continuous operators   
$W, r(t):{\cal H}'\to {\cal H}'_\de$ such that   
for $\Psi\in{\cal H}'$ 
\be\la{6.7}   
U'(t)\Psi =U'_0(t)W\Psi+r(t)\Psi,\,\,t\geq 0,   
\ee   
and for any $R>0$ there exists a constant  
$C_R=C(R,\de,\gamma)$ such that    
\be\la{6.8}   
\brr r(t)\Psi \brr'_{\delta}   
\le C_R e^{-\gamma t}   
\Vert \Psi\Vert'_{(R)} ,~~t\ge 0.   
\ee   
\end{pro}   
{\bf Proof.}   
We apply the standard Cook method: see, e.g.,  
\ci[Thm XI.4]{RS3}.   
Fix  $\Psi\in{\cal H}_{(R)}'$ and   
define $W\Psi$,  formally, as   
\be\la{6.9}   
W\Psi=\lim_{t\to\infty}U'_0(-t)U'(t)\Psi  =  U'_0(-T_1)U'(T_1) 
\Psi+   
\int\limits_{T_1}^{\infty}\frac{d}{dt}   
U'_0(-t)U'(t)\Psi  \,dt  
\ee   
with an appropriate $T_1>0$. 
We have to prove   
the convergence of the integral in the norm of the space ${\cal H}'_\de$.   
First,   observe that 
$$   
\frac{d}{dt} U'_0(t)\Psi={\cal A}'_0 U'_0(t)\Psi,~~~~~~~~   
\frac{d}{dt} U'(t)\Psi={\cal A}' U'(t)\Psi,    
$$   
where ${\cal A}'_0$ and  ${\cal A}'$   
are the generators to the groups $U'_0(t)$,  $U'(t)$,   
respectively.   
Similarly to (\ref{A0'}), we have   
\be\la{A'}   
{\cal A}'=\left(   
\begin{array}{cc}   
0 & A\\   
1 & 0   
\end{array}\right),   
\ee   
where $Au=\sum\pa_i(a_{ij}(x)\pa_ju)-a_0(x)u$.   
Therefore,   
\be\la{6.10}   
\frac{d}{dt} U'_0(-t)U'(t)\Psi   
=U'_0(-t) ({\cal A}'-{\cal A}'_0) U'(t)\Psi.   
\ee   
(\ref{A'}) and (\ref{A0'}) imply   that 
\be\la{6.11}   
{\cal A}'-{\cal A}'_0 =   
\left( \begin{array}{cc} 0 & A-A_0  \\   
0 & 0 \end{array}\right).   
\ee   
Observe  that $A-A_0=   
\sum\pa_i b_{ij}(x)\pa_j-a_0(x)$,   
where  $b_{ij}(x),$ $a_0(x)\in C_0^{\infty}(B_{R_0})$   
with some $R_0>0$,   
according to {\bf E1}. Therefore,    
 by (\ref{6.5}),   we have that  
\beqn 
&&\brr U'_0(-t) ({\cal A}'-{\cal A}'_0) U'(t)\Psi 
\brr_ \de'\le 
Ce^{\de t}   
\brr  ({\cal A}'-{\cal A}'_0) U'(t)\Psi 
\brr_ \de'\nonumber\\ 
&&\nonumber\\ 
&& 
\le Ce^{\de t}   
\Vert  \Big( ({\cal A}'-{\cal A}'_0) U'(t)\Psi\Big)^0 
\Vert_{H^{-1}(B_{R_0})} 
\le Ce^{\de t}   
\Vert  \Big(  U'(t)\Psi\Big)^1 
\Vert_{H^{1}(B_{R_0})},\,\,\,t\ge 0. 
\eeqn 
Then  (\ref{6.10})  and 
(\ref{6.6}) imply,  for $t\ge T=T(R,R_0)$,   that 
\be\la{6.13}   
\brr\frac{d}{dt} U'_0(-t)U'(t)\Psi\brr'_{\delta} \le   
C(R,R_0)   
e^{\de t} e^{-\al t}   
\Vert \Psi\Vert'_{(R)} = C(R)   
 e^{-\beta t} \Vert \Psi\Vert'_{(R)} ,   
\ee   
where $\beta=\al-\de$. Choose $\de>0$ sufficiently   
small: $\delta<\al$. Then we have $\beta>0$,   
and (\ref{6.13}) implies   
$$   
\int\limits_{T}^{+\infty}\brr\frac{d}{dt}   
U'_0(-t)U'(t)\Psi  \brr'_{\delta}\,dt\leq   
C_1(R)\Vert \Psi\Vert'_{(R)}   
<   
\infty.   
$$   
Therefore, the existence of the limit in (\ref{6.9}) follows 
if we choose $T_1=T$.   
Furthermore, the operator $W:{\cal H}'\to{\cal H}'_\de$   
is continuous, and   
  (\ref{6.9}), (\ref{6.13})  imply   
\be\la{6.14}   
\brr(U'_0(-t)U'(t)-W)\Psi\brr'_{\delta}\le   
C_2(R) e^{-\beta t} \Vert \Psi\Vert'_{(R)} ,\,\,t\geq T.   
\ee   
Let us now choose $\delta<\al/2$. Then   
$\delta<\beta=\al-\de$ and $\gamma=\beta-\delta=\al-2\delta>0$.   
Finally, 
set $r(t)\Psi:=U'(t)\Psi - U'_0(t)W\Psi$, then 
(\ref{6.14}) and (\ref{6.5}) imply   
$$\ba{rcl}   
\brr r(t)\Psi\brr'_\de&=&   
\brr(U'(t) - U'_0(t)W)\Psi\brr'_{\delta}   
=\brr U'_0(t)(U'_0(-t)U'(t) - W)\Psi\brr'_{\delta} \\   
~&&\\   
&\le&   
C_3(R)e^{\de t}   
\brr(U'_0(-t)U'(t)-W)\Psi\brr'_{\delta}\le   
C_4(R) e^{-\gamma t}   
\Vert \Psi\Vert'_{(R)} ,~~t\ge T. ~~~~~~~~~~~~~~~ \Box 
\ea   
$$   
~\medskip\\   
{\bf Proof of Theorem \ref{t6.1}}. (\ref{6.7})   
implies that for   
$Y_0\in{\cal H}$ and  $\Psi\in{\cal H}'_{(R)}$, for any $R>0$,   
\be\la{6.7d}   
\langle U(t)Y_0,\Psi\rangle =\langle U_0(t)Y_0,W\Psi\rangle  
+\langle Y_0,r(t)\Psi\rangle ,\,\,t\geq 0   
\ee   
By Proposition \ref{t6.2}, operators $W_R\equiv W|_{{\cal H}'_{(R)}}$ and   
$r_R(t)\equiv r(t)|_{{\cal H}'_{(R)}}$ are continuous  as maps  
${\cal H}'_{(R)}\to {\cal H}'_\de$.   
Therefore, the reflexivity of the Hilbert spaces  
implies the existence of   
the adjoint continuous operators   
$\Theta_R=W_R':{\cal H}_\de\to {\cal H}_{(R)}$ and   
$\rho_R(t)=r_R'(t):{\cal H}_\de\to {\cal H}_{(R)}$ for any $R>0$.   
It remains to define $(\Theta Y_0)|_{B_R}=\Theta_R Y_0$ and   
$(\rho(t) Y_0)|_{B_R}=\rho_R(t) Y_0$ for any  $R>0$.   
\hfill$\Box$

\setcounter{equation}{0}   
\section{Convergence to equilibrium for variable coefficients}   
   
We deduce Theorem A  from  next  
two    
Propositions \re{l7.1} and \re{l7.2} 
 (cf. Propositions \ref{l2.1}, \ref{l2.2}).   
   
\begin{pro}\la{l7.1}   
Family of measures $\{\mu_t,~t\in \R\}$,   
is weakly compact in   
 ${\cal H}^{-\ve },~\forall \ve >0$.   
\end{pro}   
   
\begin{pro}\la{l7.2}   
For every $\Psi\in {\cal D}$,   
\be\la{7.1}   
\hat \mu_t(\Psi )\equiv\int \exp(i\langle Y,\Psi\rangle )\mu_t(dY)   
\rightarrow \exp\{-\fr{1}{2}{\cal Q}_\infty (\Psi ,\Psi)\},   
~~t\to\infty.   
\ee   
\end{pro}   
{\bf Proof of Proposition \ref{l7.1}}.   
Similarly to  Proposition \re{l2.1}  , 
Proposition \ref{l7.1}   
follows   
from  the  bounds   
\be\la{7.2}   
\sup\limits_{t\geq 0}   
E  \Vert U(t) Y_0\Vert_{R}^2<\infty,\,\,\,\, R>0.   
\ee   
Theorem \ref{t6.1} implies that   
$$   
E  \Vert U(t) Y_0\Vert_{R}^2\le   
2 E  \Vert \Theta  U_0(t) Y_0\Vert_{R}^2 +   
2 E  \Vert r(t) Y_0\Vert_{R}^2\le   
C_1(R)E\brr U_0(t) Y_0\brr^2_{\delta} +   
C_2(R) e^{-2\gamma|t|} E\brr Y_0\brr^2_{\delta}.   
$$   
Then (\ref{7.2}) follows from (\ref{bdt}) and (\ref{bd}). 
\hfill$\Box$\\   
{\bf Proof of Proposition \ref{l7.2}}.   
Let $\Psi\in {\cal H}'_{(R)}$. Then   
Theorem \ref{t6.1} implies that   
\beqn\la{7.3}   
\hat\mu_t(\Psi)\equiv E\exp\{i\langle  U(t)Y_0,\Psi\rangle \}&=&   
E\exp\{i\langle  \Theta U_0(t) Y_0+r(t)Y_0,\Psi\rangle \}\nonumber\\ 
&=&   
E\exp\{i\langle  \Theta U_0(t) Y_0,\Psi\rangle \}+\nu(t),   
\eeqn   
where   
$$   
\nu(t)=   
E\left[   
\exp\{i\langle  \Theta U_0(t) Y_0,\Psi\rangle \}   
(\exp\{i\langle  r(t)Y_0,\Psi\rangle \}-1) \right].   
$$   
Note that $\nu(t)$ vanishes as $t\to\infty$.   
In fact,  Theorem \ref{t6.1}  implies  as above,  
\be\la{7.4}   
|\nu(t)|\le E~|\!\langle  r(t)Y_0,\Psi\rangle \!|  
\le C(R)\Vert\Psi\Vert'_{(R)}   
E\Vert r(t)Y_0\Vert_{R}   
\to 0,~~t\to\infty.   
\ee   
Finally, Proposition \re{l2.2} and Corollary \re{coH} imply that   
\be\la{7.5}   
E\exp\{i\langle  \Theta U_0(t) Y_0,\Psi\rangle \}=   
E\exp\{i\langle  U_0(t) Y_0,W\Psi\rangle \}\to   
\exp\{-\frac{1}{2}{\cal Q}_{\infty}(W\Psi,W\Psi)\},   
~~t\to\infty,   
\ee   
and 
(\ref{7.3})--(\ref{7.5}) imply (\ref{7.1}).   
\hfill$\Box$

\setcounter{equation}{0}   
\section{ Vainberg's estimates}   
In this section we   prove  Lemma \re{l6.2}.   
\begin{pro}\la{l1b}   
Let {\bf E1}--{\bf E3} hold, and  $n\ge 3$ be odd.   
 Then   
$\forall R,R_0>0$  there exist constants  
$\al, C(R,R_0)>0$ and   
$T=T(R,R_0)>0$ such that   
for  $Y_0\in {\cal H}$ with $\supp Y_0\subset B_{R_0}$, 
\be \la{b1}   
\Vert \frac{\partial ^k}{\pa t^k}U(t) Y_0\Vert_R  \le   
C(R,R_0) e^{-\al t}   
\Vert Y_0\Vert_{R_0},\quad t\ge T,\quad k=0,1,  \dots    
\ee     
\end{pro}  
{\bf Proof}~  
Conditions {\bf E1}-{\bf E3} allow us to use 
Theorem  X.4 from \ci{V89} which implies  an 
asymptotic expansion  
for the solution $Y(\cdot,t)=U(t) Y_0$, 
\be\la{aex} 
Y(x,t)=\sum\limits_{j=1}^N\sum_{l=0}^{d_j} 
t^l  
e^{-i\om_jt} 
Y_{jl}(x) 
+Z_N(x,t). 
\ee 
Here Im~$\om_j$ is a nonincreasing sequence, 
Im~$\om_j\to-\infty$,  
$d_j<\infty$, and 
  the remainder $Z_N(x,t)$ satisfies the bounds (\ref{b1}) 
when Im~$\om_{N+1}<0$. 
It remains to prove that all terms with Im~$\om_N\ge 0$ 
vanish. 
Assumptions {\bf E1} and {\bf E2}  provide an  a priori 
bound for finite energy solutions. Therefore,  
in accordance with  Theorem 8 (or Lemma 10) of \ci[Ch X]{V89}, 
all increasing terms  
with  Im~$\om_j>0$ or with   Im~$\om_j=0$ and $l>0$ 
 vanish. By the same reason, 
each 
 amplitude  
$Y_{jl}(x)$ 
 with Im$~\om_j=0$ and $l=0$ has a  finite global energy 
and 
hence  is an eigenfunction of  
the operator  $L$ (see (\re{1.1})) 
 with the eigenvalue 
$-\om_j^2$. 
 Therefore, an  extension  of Kato's Theorem 
\ci[Thm 2.1]{Ei}  
 implies the absence of a discrete spectrum inside the continuous 
spectrum, i.e. 
 $Y_{jl}(x)=0$ if $\om_j\ne 0$. 
Amplitude with  $\om_j=0$ and $l=0$ vanishes  since the operator $L$ 
is strictly negative. 
 \hfill$\Box$ 
  
Therefore, we get by   
 duality the following  bounds for   $\Psi\in {\cal H}'_{(R)}$: 
\be \la{b2}   
\Vert \frac{\partial ^k}{\pa t^k}U'(t) \Psi   
\Vert_{H^{-1}(B_{R_0})\oplus L^2(B_{R_0})} \le   
C(R,R_0) e^{-\al t}   
\Vert \Psi\Vert'_{(R)},~~t\ge T,\quad k=0,1,\dots       
\ee

Recall that   
(\ref{UP}) implies the representation   
$U'(t)\Psi=(\dot\psi(\cdot,t),\psi(\cdot,t))$    
where $\psi(x,t)$ is a solution to $\ddot \psi=A\psi$.   
Then (\ref{b2}) with $k=0$ implies   
\be\la{b3}   
\Vert \psi(\cdot,t)\Vert_{L^2(B_{R_0})}\le   
C(R,R_0) e^{-\al t}   
\Vert \Psi\Vert'_{(R)},~~t\ge T.   
\ee   
Similarly, (\ref{b2}) with $k=1$  implies that   
\be\la{b4}   
\Vert \dot\psi(\cdot,t)\Vert_{L^2(B_{R_0})}\le   
C(R,R_0) e^{-\al t}   
\Vert \Psi\Vert'_{(R)},~~t\ge T.   
\ee   
Also, by virtue of   
$\ddot \psi=A\psi$, we get   
\be\la{b5}   
\Vert A\psi(\cdot,t)\Vert_{H^{-1}(B_{R_0})}\le   
C(R,R_0) e^{-\al t}   
\Vert \Psi\Vert'_{(R)},~~t\ge T.   
\ee   
Note that   (\ref{b4}) is a part of the bound (\ref{6.6}).   
It remains to ob\-tain the bound for   
$\Vert\psi(\cdot,t)\Vert_{H^1(B_{R_0})}$.   
We deduce this bound  from the interior Schauder estimates  
\ci[Thm VI.5]{V89}   
for an elliptic operator $A$:   
\be\la{b6}   
\Vert \psi(\cdot,t)\Vert_{H^{1}(B_{R_0})}\le   
C\left(\Vert A\psi(\cdot,t)\Vert_{H^{-1}(B_{R_0+1})}+   
\Vert \psi(\cdot,t)\Vert_{L^{2}(B_{R_0+1})}\right).   
\ee   
We use (\ref{b3}) and (\ref{b5}) with $R_0+1$ instead   
of $R_0$   
in the right hand side of (\ref{b6})   
and obtain   
\be\la{b7}   
\Vert \psi(\cdot,t)\Vert_{H^{1}(B_{R_0})}\le   
C_1(R,R_0) e^{-\al t}   
\Vert \Psi\Vert'_{(R)},~~t\ge T(R,R_0+1).   
 \ee   
Therefore, (\ref{b4}), (\ref{b7})  imply (\ref{6.6}).   
\hfill$\Box$

\setcounter{equation}{0} 
\section{Appendix. Fourier transform calculations}

We consider dynamics and CFs of the solutions 
to the system (\ref{2'}).   
Let  $F:~w\mapsto\hat w$  
denote the FT of a tempered distribution 
$w\in S'(\R^n)$ (see, e.g. \ci{EKS}). 
We  also use this notation for vector- 
and matrix-valued functions. 
  
\subsection{Dynamics in the Fourier space} 
In the Fourier representation, the system (\ref{2'}) becomes 
$\dot{\hat Y}(k,t)=\hat{\cal A}_0 (k)\hat Y(k,t)$, hence  
\be\la{Frep} 
\hat Y(k,t)=\hat{\cal G}_t( k) 
\hat Y_0(k), 
\,\,\,\,\,\,\hat{\cal G}_t( k)=\exp({\hat{\cal A}_0(k)t}). 
\ee 
Here we denote  
\be\la{hatA} 
\hat{\cal A}_0(k)= 
\left( \begin{array}{ccc} 0 &~~& 1 \\ 
~\\ 
 -|k|^2 &~~& 0 \end{array}\right), 
\,\,\,\,\,\,\,\,\,\,\quad\quad 
\hat{\cal G}_t( k)= 
\left( \begin{array}{ccc} {\rm cos}~|k| 
 t &~~& \ds\fr{\sin |k| t}{|k|}  \\ 
~\\ 
 -|k|~{\rm sin}~|k| t 
&~~&  {\rm cos}~|k| t\end{array}\right). 
\ee 
 
\subsection{Cvariance functions in Fourier space} 
Translation invariance  (\ref{1.9'}) implies 
that in the sense of distributions 
\be\la{tid} 
E\Big( \hat Y_0(k)\otimes_C {\hat Y_0(k')}~\Big)=F_{x\to k}F_{y\to k'}q_0(x-y)= 
(2\pi)^n 
\de( k+k')\hat q_0( k),  
\ee 
where $\otimes_C$ stands for tensor product of complex vectors.
Now (\ref{Frep}) and (\ref{hatA}) give 
in the matrix notation, 
\be\la{tidt} 
E\Big(\hat Y(k,t)\otimes_C{\hat Y(k',t)}~\Big)= 
(2\pi)^n 
\de( k+k')\hat{\cal G}_t(k)\hat q_0( k)\hat{\cal G}'_t(k). 
\ee 
Therefore,  
\be\la{tidtx} q_t(x-y):=
E \Big(Y(x,t)\otimes_C Y(y,t)\Big)=F^{-1} 
\hat{\cal G}_t(k)\hat q_0( k)\hat{\cal G}'_t(k). 
\ee


\end{document}